\documentclass[aps,reprint,showpacs,superscriptaddress,showkeys,noeprint]{revtex4-1} 
\usepackage[utf8]{inputenc}
\usepackage{cmap} 
\usepackage[T1]{fontenc}
\usepackage{microtype}
\usepackage{amsmath,amssymb,mathtools}
\usepackage{txfonts}

\usepackage{mathrsfs}
\usepackage{graphicx,grffile,textcomp}
\usepackage{textcomp}
\usepackage{booktabs,tabularx,dcolumn}
\newcolumntype{d}[1]{D{.}{.}{#1}}

\usepackage{url,hyperref}
\usepackage[usenames,dvipsnames]{xcolor}
\hypersetup{colorlinks=true, linkcolor=BrickRed, urlcolor=blue!50!black, citecolor=blue!50!black}
\begin{document}
\title{Understanding nonequilibrium scaling laws governing collapse of a polymer}
\author{Suman Majumder}\email[]{suman.majumder@itp.uni-leipzig.de}  
\affiliation{Institut f\"ur Theoretische Physik, Universit\"at Leipzig, IPF 231101,
04081 Leipzig, Germany}
\author{Henrik Christiansen}\email[]{henrik.christiansen@itp.uni-leipzig.de}
\affiliation{Institut f\"ur Theoretische Physik, Universit\"at Leipzig, IPF 231101,
04081 Leipzig, Germany}
\author{Wolfhard Janke}\email[]{wolfhard.janke@itp.uni-leipzig.de}
\affiliation{Institut f\"ur Theoretische Physik, Universit\"at Leipzig, IPF 231101,
04081 Leipzig, Germany}
\date{\today}
%

%
\begin{abstract}
Recent emerging interest in experiments of single-polymer dynamics urge 
computational physicists to revive their understandings, particularly in the 
nonequilibrium context. Here we briefly discuss the currently evolving approaches 
of investigating the evolution dynamics of homopolymer collapse in computer simulations. Primary focus 
of these approaches is to understand various dynamic scaling laws related to 
coarsening and aging during the collapse in space dimension $d=3$, using tools popular in 
nonequilibrium coarsening dynamics of particle or spin systems. In addition to providing 
an overview of those results, we also present new preliminary data for $d=2$.  
\end{abstract}
\maketitle
\section{Introduction}\label{intro}
Understanding various scaling laws governing a phase transition has been one of the 
primary research topics over the last fifty years, be it from an equilibrium perspective or at the 
nonequilibrium front \cite{landau1958,Stanleybook,Onukibook,Puribook}.
Also for polymers, the equilibrium aspects of phase transitions have been studied extensively \cite{deGennesbook,Doibook,Cloizeauxbook,rubinstein2003polymer}.
Polymers in general represent a large class of macromolecules be they chemically synthesized or naturally occurring. 
A range of fundamentally important biomolecules, e.g., proteins and DNA, fall under the broad canopy of polymers. 
Most of these polymeric systems exhibit some form of conformational phase transitions depending on certain external 
conditions, viz., the collapse transition in homopolymers. Upon changing the solvent condition 
from good (where monomer-solvent interaction is stronger) to poor (where monomer-monomer interaction is stronger), 
a homopolymer undergoes a collapse transition from its extended coil state to a 
compact globule \cite{stockmayer1960,nishio1979}. This transition belongs to a class of phase transitions that can be understood
by investigating various associated scaling laws \cite{deGennesbook,Doibook,Cloizeauxbook,rubinstein2003polymer}. 
From a general point of view, the understanding of the collapse transition in homopolymers can be extended to investigate other 
conformational transitions experienced by different types of macromolecules, e.g., in a protein the collapse of the backbone 
may occur simultaneously or precede its folding to a native state \cite{camacho1993,Pollack2001,Sadqi2003,haran2012,reddy2017}. 
\par
Due to certain technical difficulties such as preparing a super-dilute solution or finding a long 
enough polymer with negligible polydispersity, the experimental realization of the collapse 
transition was rare in the past \cite{nishio1979,chu1995}. Since the introduction of technical 
equipment like small angle x-ray scattering, single molecule fluorescence, dynamic 
light scattering, dielectric spectroscopy, etc., monitoring the behavior of a single macromolecule 
has become feasible \cite{schuler2002,Xu2006,tress2013}. On the other hand, theoretically the scaling laws 
related to the static and the equilibrium dynamic aspects of the transition are well 
understood since a long time \cite{deGennesbook,Doibook,Cloizeauxbook,rubinstein2003polymer}. 
\par
In contrast to the equilibrium literature, however, in the nonequilibrium aspects, i.e., for the kinetics 
of the collapse transition, there is no unanimous theoretical understanding even though quite a few analytical 
and computational studies have been conducted \cite{deGennes1985,byrne1995,timoshenko1995,kuznetsov1995,kuznetsov1996,kuznetsov1996eDNA,dawson1997,pitard1998,klushin1998,Halperin2000,kikuchi2002,Abrams2002,Montesi2004,yeomans2005,pham2008,guo2011}.
The aforesaid experimental developments to track single polymers and the lack of 
understanding of the nonequilibrium dynamics of polymers motivated us to perform a series of works on the kinetics of polymer collapse
\cite{MajumderEPL,Majumder2016PRE,majumder2016proceeding,majumder2017SM,christiansen2017JCP,majumder2018proceeding}. There our novel 
approach of understanding the collapse by using its analogy with usual coarsening phenomena of particle and spin systems provided intriguing 
new insights, as will be discussed subsequently.
\par
Most of the studies on collapse kinetics in the past dealt with the understanding of the relaxation time, i.e., the time a system requires to attain its new equilibrium state once its current state is perturbed by a sudden change of the environmental conditions, e.g., the temperature. 
In the context of polymer collapse, the relaxation time is referred to as the collapse time $\tau_c$, which measures the time a polymer 
that is initially in an extended state needs to reach its collapsed globular phase. Obviously, $\tau_c$ depends on the degree of polymerization or chain length $N$ 
(the number of repeating units or monomers in the chain) of the polymer, which can be understood via the scaling relation
\begin{equation}\label{tau_scaling}
 \tau_c \sim N^z,
\end{equation}
where $z$ is the corresponding dynamic exponent. The above relation is reminiscent of the scaling one observes for dynamic 
critical phenomena \cite{HH-review}. The other important aspect of the 
kinetics is the growth of clusters of monomers that are formed during the collapse \cite{byrne1995,Abrams2002}. The 
cluster growth has recently been understood by us using the phenomenological similarities of collapse with 
coarsening phenomena in general \cite{MajumderEPL,majumder2017SM,christiansen2017JCP}. Moreover, along the same line one can also find evidence 
of aging and related scaling laws  \cite{Majumder2016PRE,majumder2016proceeding,majumder2017SM,christiansen2017JCP} that was mostly ignored in the past. 
\begin{table*}[t!]\label{tab_for_tauc}
  \caption{Summary of the simulation results for the scaling of the collapse time $\tau_c$ with the length of the polymer $N$ 
  as described in \eqref{tau_scaling}.}
  \centering
  \begin{tabular}{r c c c c c}
    \hline
    
    Authors~~~~~~~~~~~                                           & Model                              &   Method        &Explicit Solvent & Hydrodynamics & $z$  \\
    \hline
     
    Byrne \textit{et al.} (1995) \cite{byrne1995}                 & Off-lattice                        & Langevin        & No              & No  &   $3/2$\\
    Kuznetsov \textit{et al.} (1995) \cite{kuznetsov1995}         & Lattice                            & MC simulations  & No              & No &    $2$\\
    Kuznetsov \textit{et al.} (1996) \cite{kuznetsov1996}         & GSC equations                      & Numerically     & No              & No &    $2$\\
    Kuznetsov \textit{et al.} (1996) \cite{kuznetsov1996}         & GSC equations                      & Numerically     & No              & Yes &   $3/2$\\
    Kikuchi  \textit{et al.} (2005) \cite{yeomans2005}            & Off-lattice                        & MD simulations  & Yes             & No &    $1.89(9)$\\
    Kikuchi  \textit{et al.} (2005) \cite{yeomans2005}            & Off-lattice                        & MD simulations  & Yes             & Yes &   $1.40(8)$\\
    Pham \textit{et al.} (2008) \cite{pham2008}                   & Off-lattice                        & BD simulations  & No              & No &    $1.35(1)$\\
    Pham \textit{et al.} (2008) \cite{pham2008}                   & Off-lattice                        & BD simulations  & No              & Yes &   $1.01(1)$\\
    Guo \textit{et al.} (2011) \cite{guo2011}                     & Off-lattice                        &DPD simulations  & Yes             & Yes &   $0.98(9)$\\
    Majumder \textit{et al.} (2017) \cite{majumder2017SM}         & Off-lattice                        & MC simulations  & No              & No &    $1.79(6)$\\
    Christiansen \textit{et al.} (2017) \cite{christiansen2017JCP}& Lattice                            & MC simulations  & No              & No &    $1.61(5)$\\
       \hline
    
  \end{tabular}
 \end{table*}
\par
In this Colloquium, we intend to give a brief review of the results available on collapse 
kinetics based on the above mentioned three topics: relaxation, coarsening, and aging. It is organized in the following way. We 
will begin with an overview of the phenomenological theories of collapse dynamics followed by an overview of the previous simulation results in Section\ \ref{Overview_prev}. Afterwards, in Section\ \ref{recent_MC}, we will 
discuss our recent developments concerning the understanding of relaxation time, cluster growth and aging for the kinetics of the collapse transition in a homopolymer. Then we will present in Section\ \ref{2D} some preliminary results on the special case of polymer collapse kinetics in space dimension $d=2$. In Section\ \ref{conclusion}, finally, we wrap up with a discussion and an outlook to future research in this direction.
\section{Overview of previous studies on collapse dynamics}\label{Overview_prev}
The first work on the collapse dynamics dates back to 1985 when 
de Gennes proposed the phenomenological \textit{sausage} model \cite{deGennes1985}. It states that 
the collapse of a homopolymer proceeds via the formation of a sausage-like intermediate 
structure which eventually minimizes its surface energy through hydrodynamic dissipation 
and finally forms a compact globule having a spherical shape. Guided by this picture, in the 
next decade there was a series of numerical work by Dawson and co-workers 
considering both lattice and off-lattice models \cite{byrne1995,timoshenko1995,kuznetsov1995,kuznetsov1996,kuznetsov1996eDNA,dawson1997}. However, the sequence of
events obtained in their simulations differs substantially from the sausage model. Later 
in 2000, Halperin and Goldbart (HG) came up with their \textit{pearl-necklace} picture of the collapse \cite{Halperin2000}, 
consistent not only with the observations of Dawson and co-workers but also with all the 
later simulation results. According to HG the collapse of a polymer upon quenching from an extended coil 
state into the globular phase occurs in three different stages:
(i) initial stage of formation of many small nascent clusters of monomers out of the 
    density fluctuations along the chain,
(ii) growth and coarsening of the clusters by withdrawing monomers 
     from the bridges connecting the clusters until they coalesce with each other to form bigger clusters and eventually forming 
     a single cluster,
and (iii) the final stage of rearrangements of the monomers within the single cluster to form a compact globule. 
Even before the pearl-necklace picture of collapse by HG, Klushin \cite{klushin1998} independently proposed a phenomenology for the 
same picture based on similar coarsening of local clusters. It differs from the HG one as it does not consider 
the initial stage of formation of the local ordering or small nascent clusters. However, almost all the simulation 
results so far have shown evidence for the initial stage of nascent cluster formation.
\par
In addition to the above description, HG also provided time scales for each of these stages which scale with the number of monomers as 
$N^0$, $N^{1/5}$ and $N^{6/5}$, respectively. Quite obviously this scaling of the collapse time 
is dependent on the underlying dynamics of the system, i.e., on the consideration of hydrodynamic effects. 
 Klushin derived that the collapse time $\tau_c$ scales as $\tau_c \sim N^{1.6}$ in absence 
of hydrodynamics whereas the collapse is much faster in presence of hydrodynamics with the scaling $\tau_c \sim N^{0.93}$ \cite{klushin1998}. 
Similar conclusions were drawn in other theoretical and simulation studies as well. In the following subsection \ref{sub_collapse_time} we discuss some
of these numerical results on the scaling of the collapse time. 
\subsection{Earlier results on scaling of collapse time}\label{sub_collapse_time}
As mentioned the dynamic exponent $z$ in Eq.\ \eqref{tau_scaling} depends on the intrinsic 
dynamics of the system. It is thus important to notice the method and even the type 
of model one uses for the computer simulations. The available 
results can be divided into three categories: (i) Monte Carlo (MC) and Langevin simulations 
with implicit solvent effect, (ii) molecular dynamics (MD) simulations with implicit solvent effect, and  
(iii) MD simulations with explicit solvent effect. Results from MC and Langevin simulations do not incorporate 
hydrodynamics and hence only mimic diffusive dynamics. On the other hand, MD simulations with implicit solvent, depending on the 
nature of the thermostat used for controlling the temperature, can be with or without hydrodynamic effects. 
At this point we caution the reader that there is a subtle difference between solvent effects and hydrodynamic effects. Thus 
doing MD simulations with explicit solvent does not necessarily mean that the hydrodynamic modes are actively taken into account. 
Rather this depends on how one treats the momenta of the solvent particles in the simulation, e.g., it depends on the choice of thermostat 
used \cite{Frenkel_book}. This gets not only reflected in the nonequilibrium relaxation times like the collapse time but also in the 
equilibrium autocorrelation time. The few existing studies on polymer collapse using MD simulations that account for solvent 
effects by considering explicit solvent beads, thus, can also be classified on the basis of consideration of 
hydrodynamic effects. Since there is no available appropriate theory for the 
nonequilibrium relaxation time, the trend is to compare the scaling of the collapse time with the available theories of equilibrium polymer dynamics. In absence of hydrodynamic effects the dynamics is compared with Rouse scaling that 
states that in equilibrium the diffusion coefficient $D$ scales with the chain length $N$ as $D \sim N^{-1}$, which implies that the relaxation time scales as $\tau \sim N^2$ \cite{Rouse1953}. On the other hand, in presence of hydrodynamics when 
the polymer moves as a whole due to the flow field, the corresponding scaling laws are $D \sim N^{-0.6}$ and $\tau \sim N$, known 
as the Zimm scaling \cite{Zimm1956}. Both Rouse and Zimm scalings have been verified in a number of computational studies as well as in experiments. 
However, we stress that the nonequilibrium relaxation time, e.g., the collapse time $\tau_c$ does not necessary follow 
the same scaling as the equilibrium autocorrelation time \cite{janke2012monte,janke2018monte}. 
\par
In Table\ \ref{tab_for_tauc} we have summarized some of the relevant results on the scaling of the collapse time that one can find in the literature.
In the early days the simulations were done mostly by using methods that do not incorporate hydrodynamics, e.g., numerical solution of the Gaussian-self
consistent (GSC) equations, MC simulations and Langevin simulations. 
They considered models which could be either on-lattice (interacting self-avoiding walks) or off-lattice (with Lennard-Jones 
kind of interaction). The GSC approach and MC simulations (in a lattice model) provided $z$ that is in agreement 
with the Rouse scaling in equilibrium \cite{kuznetsov1995,kuznetsov1996}. Langevin simulations of an off-lattice model 
yielded $z\approx 3/2$ \cite{byrne1995} which was the value later obtained in a theory by Abrams \textit{et al.} \cite{Abrams2002}.
Kikuchi \textit{et al.} \cite{kikuchi2002} went a step further by doing MD simulations of an off-lattice model with 
explicit solvent which also allows one to tune the hydrodynamic interactions. In absence of hydrodynamics they obtained 
values of $z \approx 1.9$ close to the Rouse value of $2$ \cite{yeomans2005}. On the other hand, in presence of hydrodynamic interaction the dynamics is much 
faster with $z\approx 1.4$ \cite{yeomans2005}. This is more or less in agreement with GSC results obtained considering hydrodynamic interaction 
\cite{kuznetsov1996}. Later more simulations on polymer collapse with explicit solvent were performed. In this regard, relatively recent 
Brownian dynamics (BD) simulations with explicit solvent (hydrodynamic interaction preserved) by  Pham \textit{et al.} also provided even  
faster dynamics with $z\approx 1$ \cite{pham2008}. There exist even newer results from dissipative-particle dynamics (DPD) simulation 
that also reports $z \approx 1$ \cite{guo2011}. These results can be compared with the Zimm scaling applicable to equilibrium dynamics 
in presence of hydrodynamics. The bottom line from this literature survey is that no consensus has been achieved for the value of $z$. In our recent results on collapse dynamics from 
MC simulations a consistent value of $z$ was obtained between an off-lattice model and a lattice model with $z\approx1.7$ \cite{majumder2017SM,christiansen2017JCP}.  

\subsection{Earlier results on cluster growth}
As discussed above most of the previous studies on kinetic of the collapse transition focused on understanding the scaling of the collapse time. However, going by the phenomenological picture 
described by HG, as also observed in most of the available simulation results, the second stage of the collapse, i.e., the coalescence of the ``pearl-like'' clusters to form bigger clusters and thereby eventually 
a single globule bears resemblance to usual coarsening of particle or spin systems. The nonequilibrium phenomenon of coarsening in particle or spin systems is
well understood \cite{Puribook,bray2002} with current focus shifting towards more challenging scenarios like fluid mixtures \cite{shimizu2015,basu2017}.
Fundamentally, too, it is still developing as for example in computationally expensive long-range systems \cite{Henrik2019,Janke2019CS,corberi2019}.
\par
In usual coarsening phenomena, e.g., in ordering of ferromagnets after quenching from the high-temperature disordered 
phase to a temperature below the critical point, the nonequilibrium pathway is described by a growing length scale, i.e., average linear 
size of the domains $\ell(t)$ as \cite{Puribook,bray2002} 
\begin{equation}\label{length-scaling}
 \ell(t) \sim t^{\alpha}.
\end{equation}
The value of the growth exponent $\alpha$ depends on the concerned system as well as the conservation of the order parameter during the entire process. 
For example, in solid binary mixtures where the dynamics is conserved, $\alpha=1/3$ which is the Lifshitz-Slyozov (LS) growth 
exponent \cite{LS-growth}, whereas for a ferromagnetic ordering where the order parameter 
is not conserved, $\alpha=1/2$ which is referred to as the Lifshitz-Cahn-Allen (LCA) growth \cite{LCA-growth}. On the other hand, in fluids where 
in simulations one must incorporate hydrodynamics, one observes three different 
regimes; the early-time diffusive growth where $\alpha=1/3$ as in solids; the intermediate viscous hydrodynamic growth with $\alpha=1$ \cite{Siggia1979}; and 
at a very late stage the inertial growth with $\alpha=2/3$ \cite{Furukawa1988}.
\par
In the context of polymer collapse, the concerned growing length scale 
could be the linear size (or radius) of the clusters. However, in all the previous works it was chosen to be the average mass 
$C_s(t)$, or average number of monomers present in a cluster. In spatial dimension $d$, it is related to the linear size of the cluster 
as $C_s(t) \sim \ell(t)^d$. Thus in analogy with the power-law scaling\ \eqref{length-scaling} of the length scale during coarsening, the corresponding scaling 
of the cluster growth can then be written as  
\begin{eqnarray}\label{Cs_powerlaw}
C_s(t) \sim t^{\alpha_c},
\end{eqnarray}
where $\alpha_c=d\alpha$ is the corresponding growth exponent. Like the dynamic exponent $z$, the growth exponent $\alpha_c$ is also dependent 
on the intrinsic dynamics of the system. Previous studies based on MC simulations of a lattice polymer model
reported $\alpha_{c}=1/2$ \cite{kuznetsov1995} and Langevin simulations of an off-lattice model
reported $\alpha_{c}=2/3$ \cite{byrne1995}, both being much smaller than $\alpha_c=1$ as observed for coarsening 
with only diffusive dynamics. BD simulations with explicit solvent also provided 
$\alpha_c \approx 2/3$ in absence of hydrodynamics. Like in coarsening of fluids, the dynamics of cluster growth during collapse, too, gets faster 
when hydrodynamic effects are present. For instance, BD and DPD simulations with incorporation of hydrodynamic effects yield $\alpha_c \approx 1$ \cite{guo2011,pham2008}. 
Surprisingly, our recent result on an off-lattice model via MC simulations also showed $\alpha_c \approx 1$ \cite{majumder2017SM}. This will be discussed in Section\ \ref{coarsening}.

\subsection{Earlier results on aging during collapse}
Apart from the scaling of the growth of the average domain size during a coarsening process there is another 
important aspect, namely, aging \cite{henkelbook,Zannetti_book}. The fact that a younger system relaxes faster than an older one forms the 
foundation of aging in general. This is also an essential concept from the point of view of glassy dynamics \cite{Bouchaud_book,Castillo2002}. Generally, aging is probed 
by the autocorrelation function of a local observable $O_i$ given as  

\begin{eqnarray}\label{auto_cor}
C(t,t_w)=\langle O_i(t)O_i(t_w) \rangle - \langle O_i(t) \rangle \langle O_i(t_w) \rangle, 
\end{eqnarray}
with $t$ and  $t_w < t$ being the observation and the waiting times, respectively. 
The $\langle \dots \rangle$ denotes averaging over several randomly chosen realizations of the initial configuration and 
independent time evolutions. The observable $O_i$ is generally chosen in such a way that it clearly reflects the changes happening during the concerned nonequilibrium process, e.g., 
the time- and space-dependent order parameter during ferromagnetic ordering. 
\par
There are three necessary conditions for aging: (i) absence of time-translation invariance in $C(t,t_w)$, (ii) slow 
relaxation, i.e., the relaxation times obtained from the decay of $C(t,t_w)$ should increase as function of $t_w$, and 
(iii) the observation of dynamical scaling of the form  
\begin{eqnarray}\label{autocorr_scaling}
 C(t,t_w) \sim x_c^{-\lambda}, 
\end{eqnarray}
where $x_c$ is the appropriate scaling variable and $\lambda$ is the corresponding aging or autocorrelation exponent. 
For coarsening, the scaling variable is usually taken as $x_c=t/t_w$, the ratio of the times $t$ and $t_w$, or $x_c=\ell/\ell_w$, the ratio of the corresponding growing length scales at those times.
Fisher and Huse (FH) in their study of ordering spin glasses proposed a bound on $\lambda$ which only depends on the dimension $d$ as \cite{Fisher1988}
\begin{eqnarray}\label{FH-bound}
  \frac{d}{2} \le \lambda \le d. 
\end{eqnarray}
Later this bound was found to be obeyed in the ferromagnetic ordering as well \cite{Liu1991,Lorenz2007,Midya2014}.
An even stricter and more general bound was later proposed by Yeung \textit{et al.} \cite{yeung1996} that also includes the case of conserved 
order-parameter dynamics.

\par
In the context of polymer collapse, although analogous to coarsening phenomena in general, this particular aspect of aging 
has received very rare attention \cite{pitard2001,Stanley2002}. There, like in other soft-matter systems \cite{Cloitre2000,bursac2005,wang2006} the results indicated presence of subaging, i.e., evidence for scaling similar to Eq.\ \eqref{autocorr_scaling} 
but as a function  of $x_c=t /t_{w}^{\mu}$ with $\mu <1$. Afterwards, there were no attempts to quantify this scaling with respect to the 
ratio of the growing length scale. In our approach, both with off-lattice and lattice models 
we showed that simple aging scaling as in Eq.\ \eqref{autocorr_scaling} with respect to the ratio of the cluster sizes can be observed 
\cite{Majumder2016PRE,majumder2016proceeding,majumder2017SM,christiansen2017JCP}. Thus to quantify the aging scaling, by choosing 
$x_c=C_s(t)/C_s(t_w)$ one has to transform Eq.\ \eqref{autocorr_scaling} to
\begin{eqnarray}\label{power-law_Cst}
C(t,t_w) \sim \left[\frac{C_s(t)}{C_s(t_w)}\right ]^{-\lambda_c}
\end{eqnarray}
where $\lambda_c$ is the associated autocorrelation exponent which is related to the traditional exponent $\lambda$ via the relation $\lambda_c=\lambda/d$. 
\begin{figure*}[htb!]
\centering
\resizebox{0.73\textwidth}{!}{\includegraphics{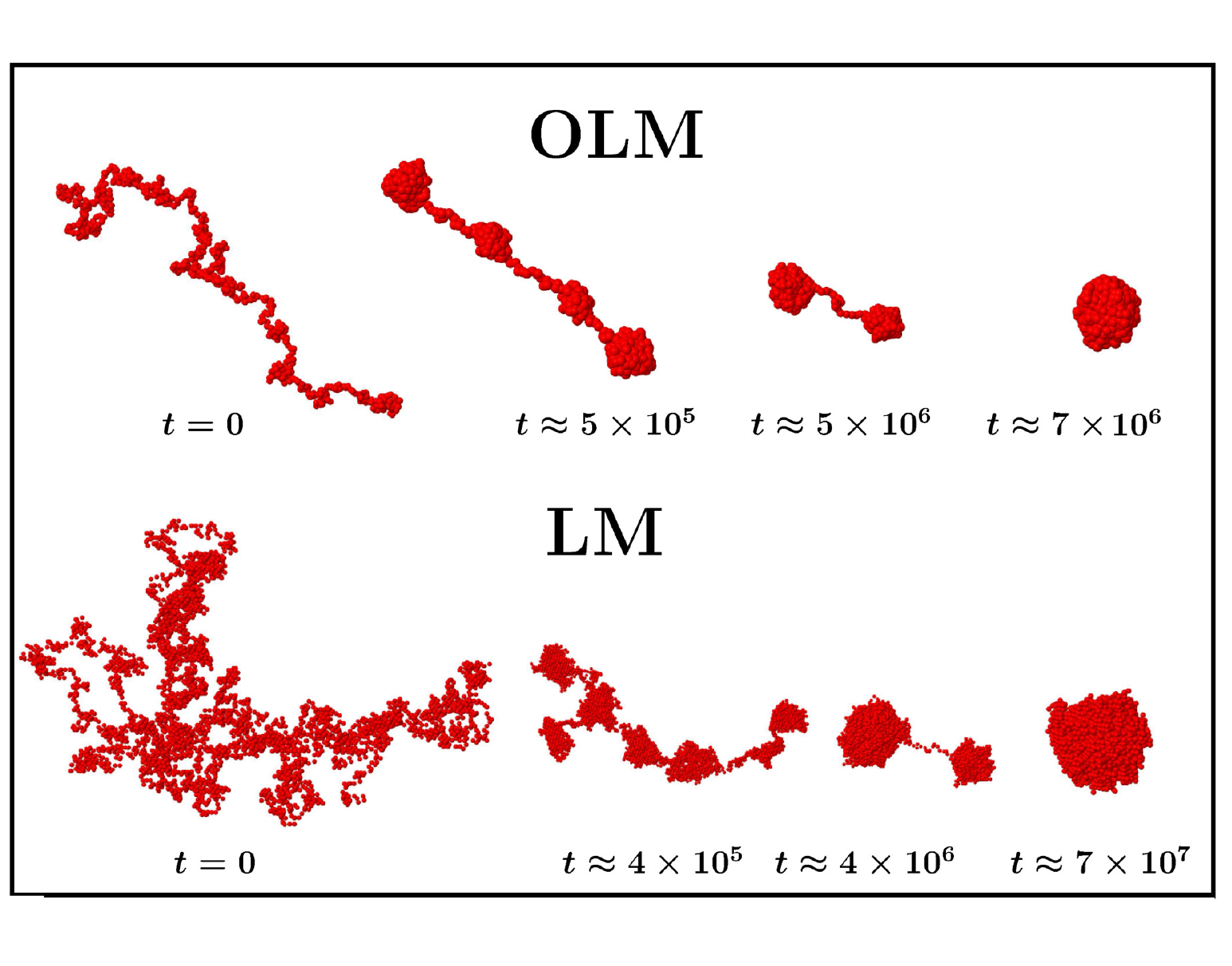}}
\caption{Time-evolution snapshots during collapse of a homopolymer showing pearl-necklace formation,
following a quench from an extended coil phase to a temperature, $T_q=1$ for OLM and $T_q=2.5$ for LM, in the globular phase. The chain lengths $N$ used are $724$ and $4096$ for OLM and LM, respectively. 
Taken from Ref.\ \cite{majumder2018proceeding}.}
\label{figsnap}      
\end{figure*}
\section{Recent Monte Carlo results in $d=3$}\label{recent_MC}
In this section we will review the very recent developments by us concerning the kinetics of homopolymer collapse from all above mentioned three perspectives. We will compare the results from an off-lattice model (OLM) and a lattice model (LM), focusing in this section on $d=3$ dimensions.
New results for the special case of $d=2$ will be presented in the next section to check the validity of the observations in general. Before moving on to a discussion of our findings next we first briefly describe the different models and methodologies used in our studies.
\subsection{Models and methods}\label{model}
For OLM, we consider a flexible bead-spring model where the connectivity between two successive 
monomers or beads are maintained via the standard 
finitely extensible non-linear elastic (FENE) potential 
\begin{eqnarray}\label{FENE}
E_{\rm{FENE}}(r_{ii+1})=-\frac{K}{2}R^2\ln\left[1-\left(\frac{r_{ii+1}-r_0}{R}\right)^2\right].
\end{eqnarray}
We chose the force constant of the spring $K=40$, the mean bond length $r_0=0.7$ and the maximum allowed deviation from the mean position $R=0.3$ \cite{milchev2001}. 
Monomers were considered to be spherical beads with diameter $\sigma =r_0/2^{1/6}$. The nonbonded interaction between the monomers 
is given by 
\begin{eqnarray}\label{potential_OLM}
E_{\rm {nb}}(r_{ij})=E_{\rm {LJ}}\left({\rm{min}}[r_{ij},r_c]\right)-E_{\rm {LJ}}(r_c), 
\end{eqnarray}
where 
\begin{eqnarray}\label{std_LJ}
E_{\rm {LJ}}(r)=4\epsilon \left[ \left( \frac{\sigma}{r} \right)^{12} - \left (\frac{ \sigma}{r} \right )^{6} \right]
\end{eqnarray}
is the standard Lennard-Jones (LJ) potential. Here $\epsilon(=1)$ is the interaction strength and $r_c$ $=2.5\sigma$ the cut-off radius.
\par
For LM, we consider a variant of the interactive self-avoiding walk on a simple-cubic lattice, where each lattice site can be 
occupied by a single monomer. The Hamiltonian is given by 
\begin{equation}\label{hamiltonian}
H=-\frac{1}{2} \sum_{i \ne j,  j \pm 1} w(r_{ij}),~~ \textrm{where}~~  
w(r_{ij})=\begin{cases} J & r_{ij} = 1 \\ 0 & \text{else}\end{cases}.
\end{equation}
Here $r_{ij}$ is the distance between two nonbonded monomers $i$ and $j$, 
$w(r_{ij})$ is an interaction parameter that considers only nearest neighbors, and $J(=1)$ is the interaction strength. 
We allowed a fluctuation in the bond length by considering diagonal bonds, i.e., the possible bond 
lengths are $1$, $\sqrt{2}$ and $\sqrt{3}$. The model has been independently studied for equilibrium properties \cite{shaffer1994,dotera1996}. 
It has certain similarities with the bond-fluctuation model \cite{carmesin1988}. For a comparison between them, please see Ref.\ \cite{subramanian2008}.
\par
The dynamics in the models can be introduced via Markov chain MC simulations \cite{janke2012monte,Landaubook}, however, with the restriction of allowing only local moves. For OLM the 
local moves correspond to shifting of a randomly selected monomer to a new position randomly chosen within [$-\sigma/10 :\sigma/10$] of its current position.
For LM, too, the move set consists of just shifting a randomly chosen mon\-omer to another lattice site such that the bond connectivity constraint is maintained.
These moves are then accepted or rejected following the Metropolis algorithm with Boltzmann criterion \cite{janke2012monte,Landaubook}. 
The time scale of the simulations is one MC sweep (MCS) which consists of $N$ (where $N$ is the number of monomers in the chain) such attempted moves. 
\par
The collapse transition temperature is $T_{\theta}(N \rightarrow \infty) \approx 2.65 \epsilon/k_B$ 
and $\approx 4.0J/k_B$ for OLM and LM, respectively \cite{majumder2017SM,christiansen2017JCP}. 
In all the subsequent discussion, the unit of temperature will always be $\epsilon/k_B$ or $J/k_B$ with 
the Boltzmann constant $k_B$ being set to unity. Following the standard protocol of nonequilibrium studies 
we first prepared initial conformations of the polymers at high temperatures $T_h \approx 1.5T_{\theta}$ 
that mimics an extended coil phase. Then this high-temperature conformation was quenched to a temperature $T_q < T_{\theta}$. Since LM is computationally less 
expensive than OLM, the chain length of polymer used for LM is longer than what is used for OLM. Note that except for the evolution snapshots, for both models, all the results 
presented were obtained after averaging over more than $300$ independent runs. For each such run, the starting conformation is an extended coil which were obtained 
independently of each other by generating self-avoiding walks using different random seeds and then equilibrating them at high temperature.

\subsection{Phenomenological picture of the collapse}
As mentioned before even though the sausage picture of de Gennes \cite{deGennes1985} is the pioneer in describing the 
phenomenology of the collapse dynamics, all simulation studies provided evidence in support of the 
pearl-necklace picture of HG \cite{Halperin2000}. In our simulations, too, both with OLM and LM, we observed intermediates that support the  
pearl-necklace phenomenology. Typical snapshots which we obtained from our simulations are shown 
in Fig.\ \ref{figsnap}. The typical sequence of events happening during the collapse are captured by these snapshots. 
At initial time the polymer is in an extended state with fluctuation of the local monomer density along the chain. 
Soon there appear a number of local clusters of monomers which then start to grow by withdrawing monomers 
from the rest of the chain. This gives rise to the formation of the so called pearl-necklace. 
Once the tension in the chain is at maximum, two successive clusters along the chain coalesce with each other to grow 
in size. This process goes on until a single cluster or globule is formed. The final stage 
of the collapse is the rearrangement of the monomers within the single cluster to form 
a compact globule. This last stage, however, is difficult to disentangle from the previous stages. 
\begin{figure*}[t!]
\centering
\resizebox{0.72\textwidth}{!}{\includegraphics{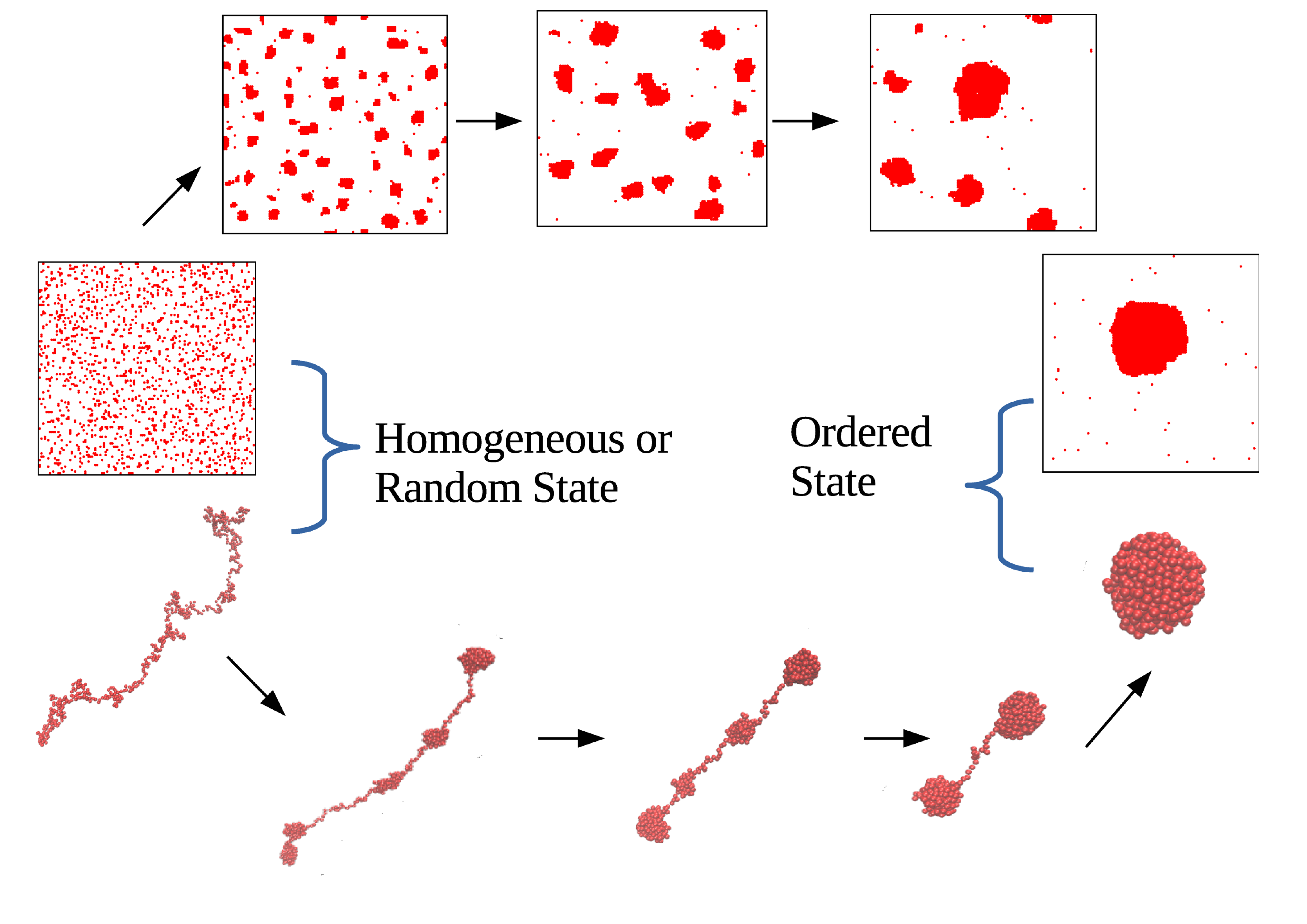}}
\caption{The upper panel shows evolution snapshots for the droplet formation in a particle system using the Ising lattice gas in two spatial dimensions. The lower panel 
shows the evolution of a homopolymer obtained from simulation of the OLM. The figure illustrates the 
similarities between the collapse kinetics with the usual coarsening of a particle system.}
\label{figschematic}      
\end{figure*}
\par
The first stages of formation and growth of clusters during the collapse of a polymer as demonstrated in Fig.\ \ref{figsnap} is clearly 
reminiscent of usual coarsening phenomena in particle or spin systems. As already mentioned traditionally for studying coarsening 
one starts with an initial state where the distribution of particles or spins is 
homogeneous, e.g., homogeneous fluid or paramagnet above the critical temperature. 
Similarly to study the collapse kinetics one starts with a polymer in an extended coil phase which is analogous to the 
homogeneous phase in particle or spin systems. Usual coarsening sets 
in when the initial homogeneous configuration is suddenly brought down to a temperature below the critical temperature where the 
equilibrium state is an ordered state, e.g., condensed droplet in fluid background or ferromagnet. Similarly, for a polymer, the collapse 
occurs when the temperature is suddenly brought down below the corresponding collapse transition temperature. 
There the equilibrium collapsed phase is analogous to the droplet phase in fluids. 
\par
Now coarsening refers to the process via which the initial homogeneous system evolves while approaching the ordered phase. 
This happens via the formation and subsequent growth of domains of like particles or spins. This is illustrated in the upper panel of 
Fig.\ \ref{figschematic} where we show the time evolution of the droplet formation in a fluid starting from a homogeneous phase via MC 
simulations of the Ising lattice gas. At early times many small domains or droplets are formed which then coarsen to form bigger droplets and eventually 
giving rise to a single domain or droplet. A similar sequence of events is also observed during collapse of a polymer as shown once again 
in the lower panel of Fig.\ \ref{figschematic} which explains the phenomenological analogy of collapse with usual coarsening phenomena. 
Coarsening from a theoretical point of view is understood as a scaling phenomenon which means that certain morphology-characterizing functions
of the system at different times can be scaled onto each other using corresponding scaling functions \cite{Puribook,bray2002}. 
This scaling in turn also implies that there must be scaling of the time-dependent length scale, too, which in most of the cases shows 
a power-law scaling like in Eq.\ \eqref{length-scaling}. Based on this understanding in general and the above mentioned analogy 
we will discuss in the remaining part of this section how to investigate the presence of nonequilibrium scaling laws in 
the dynamics of collapse of a homopolymer. 
\begin{figure}[t!]
\centering
\resizebox{0.95\columnwidth}{!}{\includegraphics{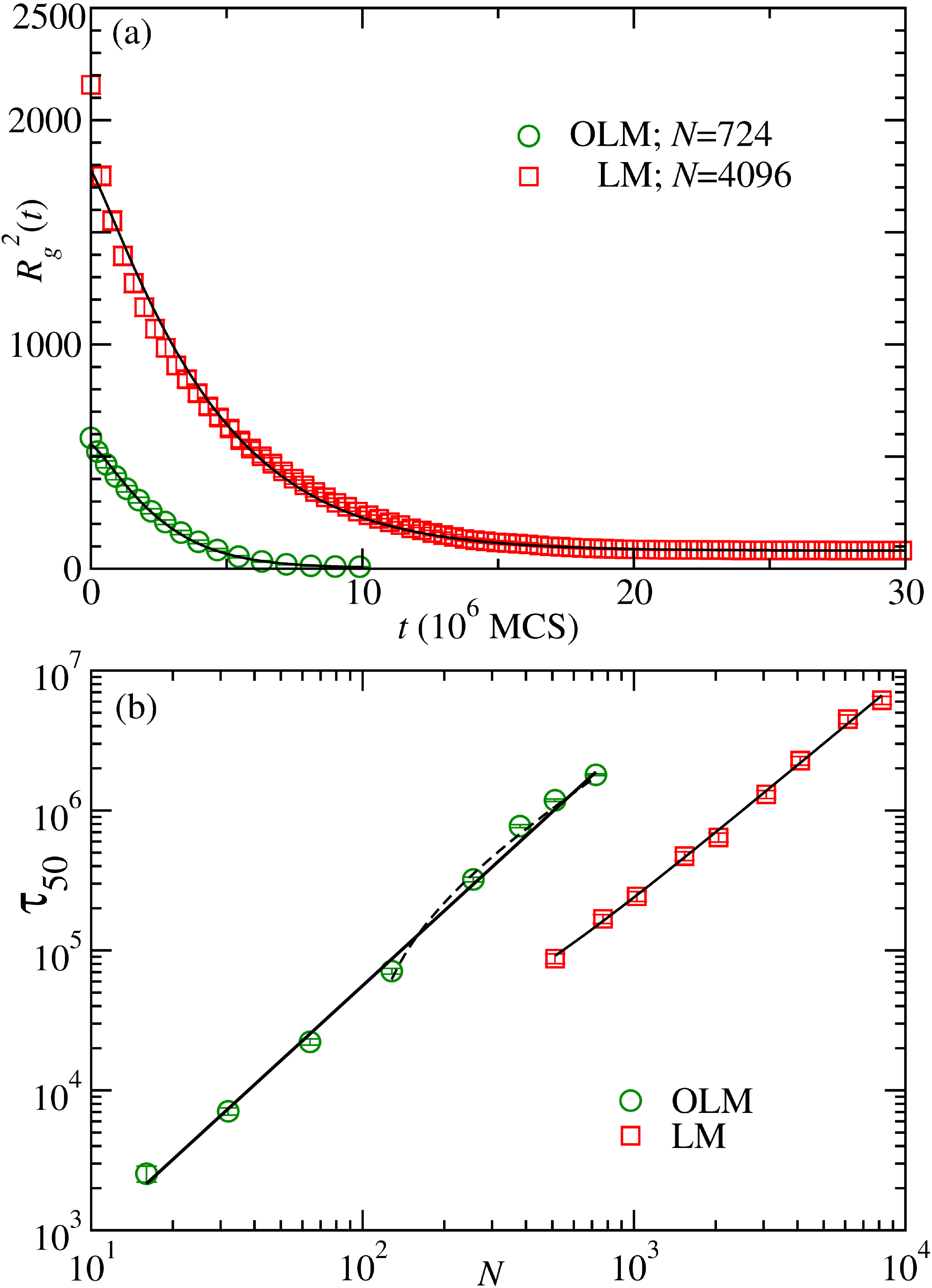}}
\caption{(a) Time dependence of the squared radius of gyration, $R_g^2(t)$, for both OLM ($T_q=1.0$) and LM ($T_q=2.5$). The solid lines are fits to a 
stretched exponential form described by Eq.\ \eqref{rg_decay}. (b) Scaling of the collapse time, $\tau_{50}$, with respect to $N$. 
The solid lines are fits to the form \eqref{tau_N}. The dashed line is a fit of the OLM data for $N\ge128$, 
to the form \eqref{tau_N} by fixing $z=1$. Adapted from Ref.\ \cite{majumder2018proceeding}.}
\label{figsrgcomp}      
\end{figure}
\subsection{Relaxation behavior of the collapse}\label{relaxation}
In all earlier studies, the straightforward way to quantify the kinetics was to monitor the 
time evolution of the overall size of the polymer, i.e., the squared radius of gyration given as 
\begin{eqnarray}\label{r_gyr}
R_{g}^2=\frac{1}{N}\sum\limits_{i=1}^{N}(\boldsymbol{r}_i-\boldsymbol{r}_{\rm{cm}})^2
\end{eqnarray}
where $\boldsymbol{r}_{\rm{cm}}$ is the center of mass of the polymer. In the coiled state (above $T_{\theta}$), $R_{g}^2 \sim N^{2\nu_{F}}$ with $\nu_{F}=3/5$, in the Flory mean-field 
approximation, whereas in the globular state (below $T_{\theta}$), $R_{g}^{2} \sim N^{2/d}$ \cite{Florybook}. Such decay of $R_{g}^2$ is shown in Fig.\ \ref{figsrgcomp}(a) for both OLM and LM. 
Although in some of the earlier studies a power-law decay of $R_{g}^2$ is suggested, in most cases or at least in the present cases that does not work. Rather, the decay can be well 
described by the form 
\begin{eqnarray}\label{rg_decay}
 R_{g}^2(t)=b_0+b_1\exp\left[-\left(\frac{t}{\tau_f} \right)^{\beta}\right],
\end{eqnarray}
where $b_0$ corresponds to the saturated value of $R_{g}^2(t)$ in the collapsed state, 
$b_1$ is associated with the value at $t=0$, and $\beta$ and $\tau_f$ are fitting parameters. 
For details about fitting the data with the form \eqref{rg_decay}, see Refs.\ \cite{majumder2017SM}  and \cite{christiansen2017JCP} for OLM and LM, respectively. 
An illustration of how appropriately this form works is shown in Fig.\ \ref{figsrgcomp}(a). There the respective solid lines are fits to the form \eqref{rg_decay}.
While the above form does not provide any detail about the specificity of the collapse process, 
it gives a measure of the collapse time $\tau_c$ via $\tau_f$.
However, to avoid the unreliable extraction of the collapse time from such a fitting, one could alternatively use a rather direct way of 
estimating $\tau_{50}$ which corresponds to the time when $R_g^2(t)$ has 
decayed to half of its total decay, i.e., $\left[R_g^2(0)-R_g^2(\infty)\right]/2$. Data for both models as shown in Fig.\ \ref{figsrgcomp}(b) reflect a power-law scaling, to be quantified with the form 
\begin{eqnarray}\label{tau_N}
\tau_c=B N^z+\tau_0,
\end{eqnarray}
where $B$ is a nontrivial constant that depends on the quench temperature $T_q$, $z$ is the 
corresponding dynamic critical exponent, and the offset $\tau_0$ comes from finite-size corrections. 
For LM a fitting (shown by the corresponding solid line) with the form \eqref{tau_N} provides $z=1.61(5)$ 
and is almost insensitive to the chosen range. However, for 
OLM the fitting is sensitive to the chosen range. While using the whole range of data provides $z=1.79(6)$ (shown by the 
corresponding solid line), fitting only the data for $N \ge128$ yields $z =1.20(9)$. In this regard, 
a linear fit [fixing $z=1$ in \eqref{tau_N}], shown by the dashed line, also works quite well. For a comparison of the values of $z$ obtained 
by us with the ones obtained by others, see Table\ \ref{tab_for_tauc}. 

\subsection{Coarsening during collapse}\label{coarsening}
Having the phenomenological analogy between collapse of a polymer and usual coarsening of particle and spin systems established, in this subsection we present the scaling of the cluster growth 
during the collapse under the light of well established protocols of the coarsening in particle or spin systems. 

\subsubsection{Scaling of morphology-characterizing functions}
\begin{figure}[t!]
\centering
\resizebox{0.92\columnwidth}{!}{\includegraphics{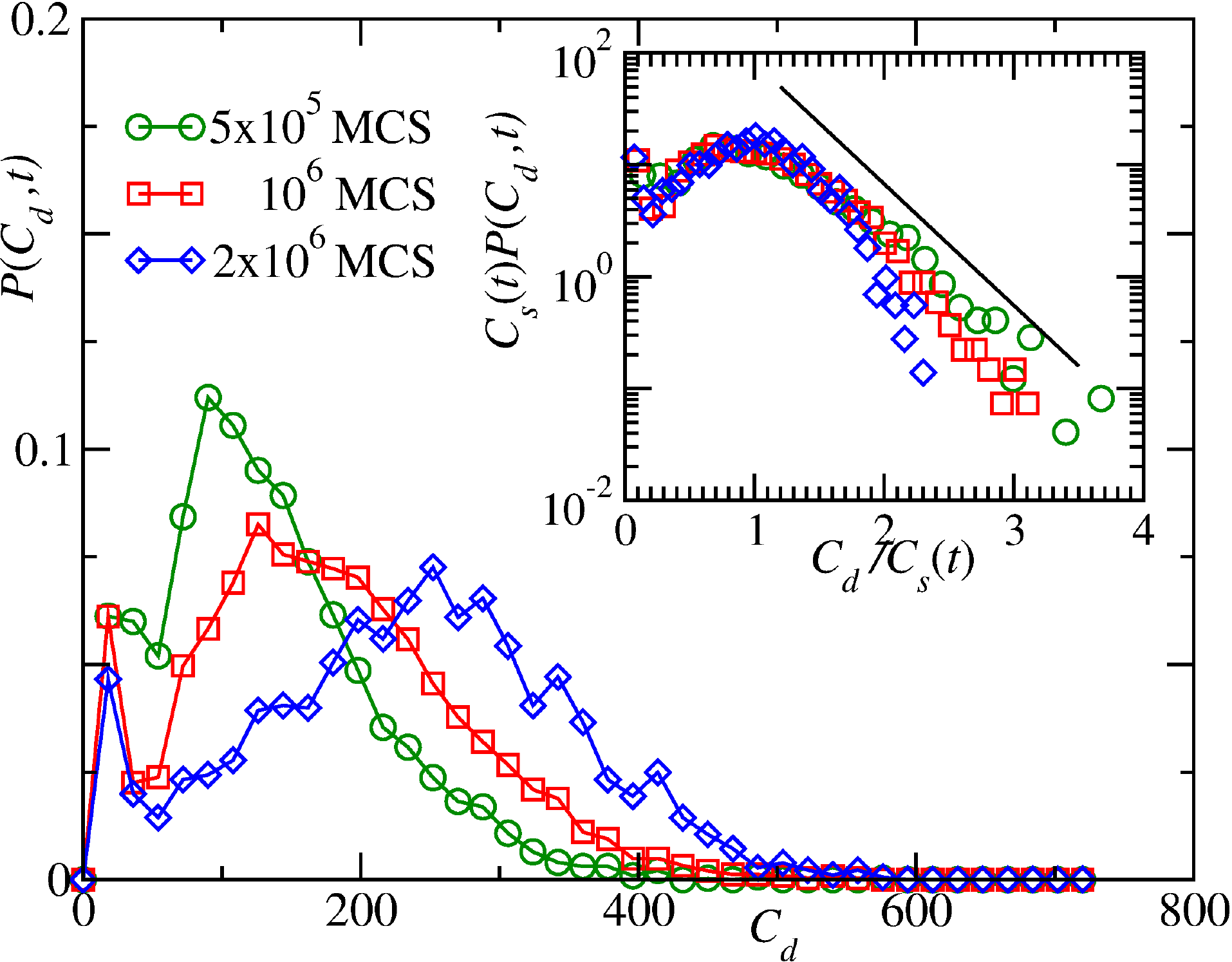}}
\caption{ Normalized distribution of the cluster sizes at three different times during the 
coarsening stage of the collapse at $T_q=1$ for a polymer with $N=724$ modeled by OLM. The inset demonstrates 
the scaling behavior of the collapse phenomenon via
the semi-log plot of the corresponding scaling of the distribution functions.  The solid line shows 
consistency of the data with an exponential tail. Taken from Ref.\ \cite{majumder2017SM}.}
\label{cdist}      
\end{figure}

Coarsening in general is a scaling phenomenon, where certain structural quantities that quantify the morphology of the system, e.g., 
two-point equal-time correlation functions and structure factors show scaling behavior \cite{Puribook,bray2002}. 
This means that the structure factors at two different times can be collapsed onto the same master curve by using the relevant length scales, i.e., cluster size or 
domain size at those times. This fact is used to extract the relevant time-dependent length scale that governs the kinetics of coarsening. 
For example one uses the first moment of the structure factor at a particular time to have a measure of the length scale or the average domain size during 
coarsening. However, to understand the kinetics of cluster growth during the collapse of a polymer traditionally the average number of monomers present 
in a cluster is used as the relevant length scale $C_s(t)$. For studying the OLM we used this definition to calculate $C_s(t)$, details of which can be found in 
Ref.\ \cite{majumder2017SM} and later will also be discussed in the $d=2$ case. The validity of this definition as the relevant length scale can be verified by 
looking at the expected scaling of the cluster-size distribution $P(C_d,t)$, i.e., the probability to find 
a cluster of size $C_d$ among all the clusters at time $t$. Using this distribution we calculate the average cluster size 
as $C_s(t)= \langle C_d \rangle $. The corresponding scaling behavior is given as 
\begin{eqnarray}\label{cdist_scaling}
 P(C_d,t) \equiv C_s(t)^{-1}  \tilde{P} [C_d/C_s(t)],
\end{eqnarray}
where $\tilde{P}$ is the scaling or master function. This means that when $C_s(t)P(C_d,t)$ at different times are plotted against $C_d/C_s(t)$ they should fall 
on top of each other. This verification is presented in Fig.\ \ref{cdist} where in the main frame we show plots of the (unscaled) distributions $P(C_d,t)$ at different times, 
and in the inset the corresponding scaling plot using the form \eqref{cdist_scaling}. Coincidentally, here, the tail of the distribution shows an exponential decay 
as observed in coarsening of particle \cite{majumder2011EPL} and spin systems \cite{Majumder2011,majumder2018_potts}.

\par
\begin{figure}[t!]
\centering
\resizebox{0.95\columnwidth}{!}{\includegraphics{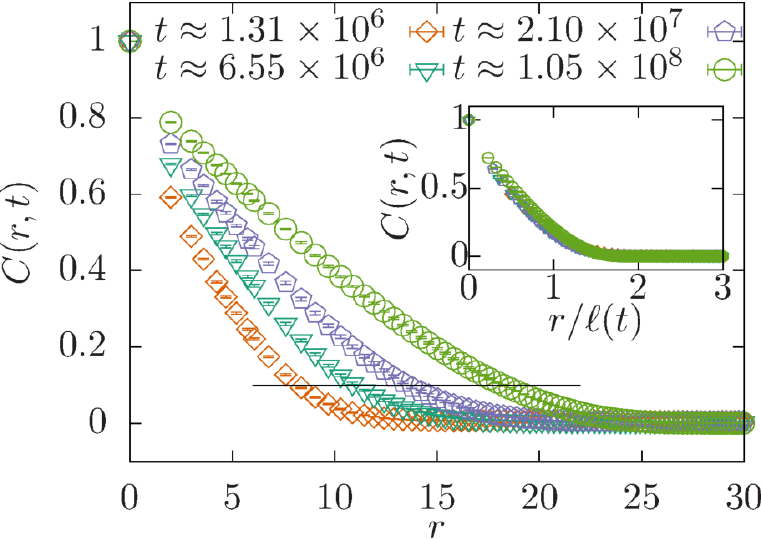}}
\caption{ Morphology characterizing two-point equal-time correlation function $C(r,t)$ at different times, showing the presence of a growing length scale 
during collapse of a polymer obtained via simulation of LM with $T_q=2.5$ and $N=8192$. The inset shows the presence of scaling in the process via the plot of the same data as a function of $r/\ell(t)$ 
where $\ell(t)$ is the characteristic length scale calculated using \eqref{lenght_from_corr} with $h=0.1$. Adapted from Ref.\ \cite{christiansen2017JCP}.}
\label{corr_LM}      
\end{figure}

On the other hand, for a lattice model, one can use the advantage of having the monomers placed on lattice points. 
There a two-point equal-time correlation function can be defined as 
\begin{eqnarray}\label{cor_lattice}
 C(r,t)=\langle \rho(0,t)\rho(r,t) \rangle
\end{eqnarray}
with
\begin{equation}\label{rho_lattice} 
\rho_i(r,t)=\frac{1}{m_r}\sum_{j, r_{ij}=r} \theta(\boldsymbol{r}_j,t).
\end{equation}
where the characteristic function $\theta$ is unity if there is a monomer at position 
 $\boldsymbol{r}_j$ or zero otherwise. $m_r$ denotes the number of possible lattice points at distance $r$ from an arbitrary point of the lattice.  
Plots for such correlation functions at different times during the collapse of a polymer using LM is shown in the main frame of Fig.\ \ref{corr_LM}. 
Slower decay of $C(r,t)$ as time increases suggests the presence of a growing length scale. Thus following the trend in usual coarsening studies one can extract 
an average length scale $\ell(t)$ that characterizes the clustering during the collapse, via the criterion 
\begin{equation}\label{lenght_from_corr}
C\left(r=\ell(t),t \right)=h,
\end{equation} 
where $h$ denotes an arbitrary but reasonably chosen value from the decay of $C(r,t)$. Calculation of $\ell(t)$ in the above manner automatically 
suggests to look for the dynamical scaling of the form
\begin{equation}
C(r,t) \equiv \tilde{C}\left(r/\ell(t)\right),
\label{EqEquiv}
\end{equation}
where $\tilde{C}$ is the scaling function.
Such a scaling behavior is nicely demonstrated in the inset of Fig.\ \ref{corr_LM}, where we show the corresponding data presented in the 
main frame as function of $r/\ell(t)$. Note that here $\ell(t)$ gives the linear size of the ordering clusters. Thus in order to compare $\ell(t)$ of LM 
with the cluster size $C_s(t)$ obtained for OLM one must use the relation $\ell(t)^d\equiv C_s(t)$. For a check of the validity of 
this relation, see Ref.\ \cite{christiansen2017JCP}.
\begin{figure}[t!]
\centering
\resizebox{0.9\columnwidth}{!}{\includegraphics{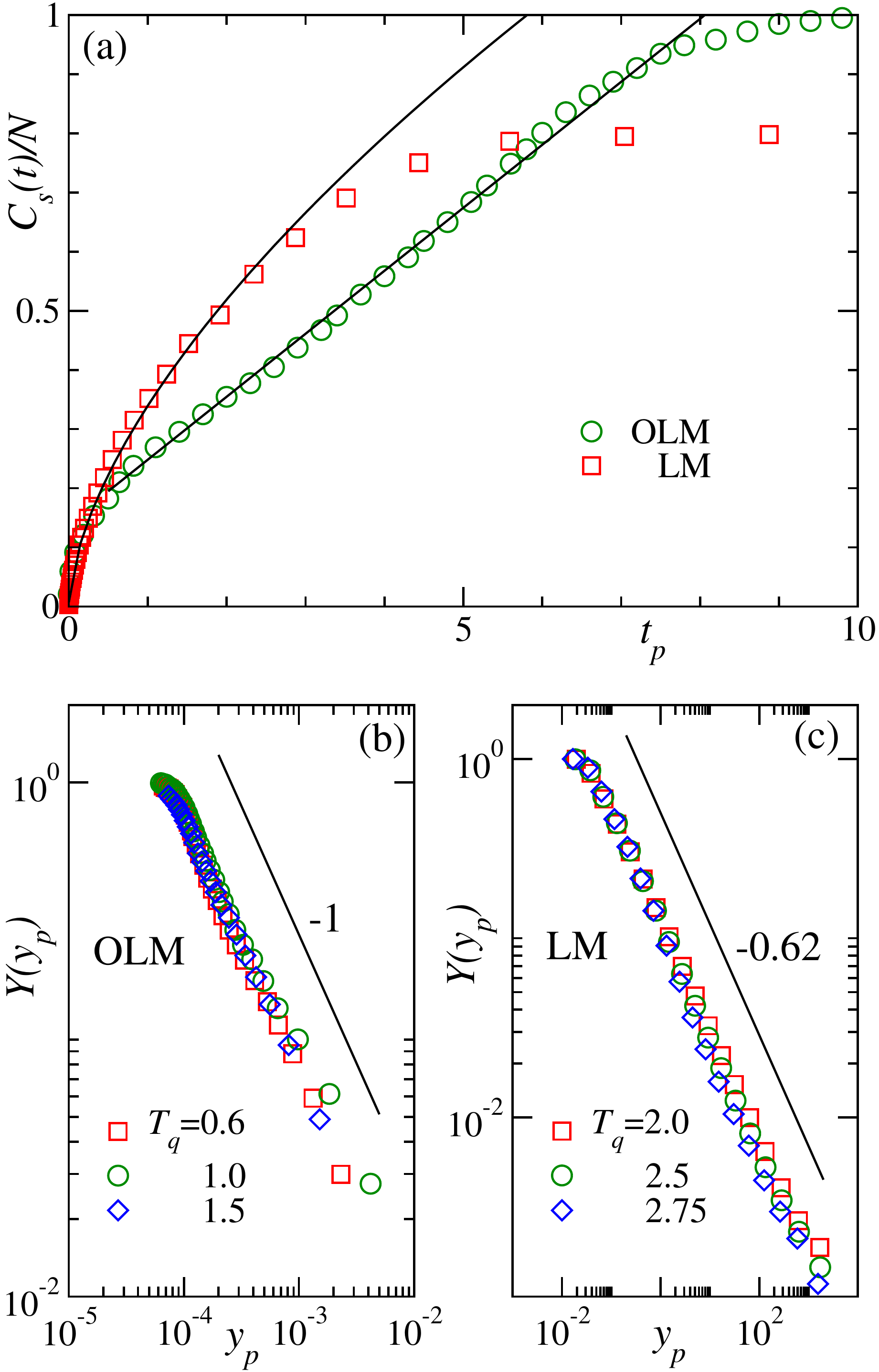}}
\caption{ (a) Plots of the average cluster size $C_s(t)/N$, as function of time for the systems 
presented in Fig.\ \ref{figsnap}. To make both the data visible on the same plot, we divide the time axis by 
a factor $m$ to obtain $t_p=t/m$, where $m=1\times10^{6}$ and $3.5\times 10^{6}$ for OLM and LM, respectively. 
The solid lines are fits to the form \eqref{cl_growth} with $\alpha_c=0.98$ for OLM and $\alpha_c=0.62$ for LM. 
The plots in (b) and (c) demonstrate the scaling exercise for OLM with $\alpha_c=1.0$ and LM with 
$\alpha_c=0.62$, respectively, showing that data for $C_s(t)$ at different quench temperatures $T_q$ 
can be collapsed onto a master curve using a nonuniversal metric factor in the scaling variable. 
The solid lines represent the corresponding $Y(y_p) \sim y_p^{-\alpha_c}$ behavior. Taken from Ref.\ \cite{majumder2018proceeding}.}
\label{cluster_size}      
\end{figure}
\subsubsection{Cluster growth}
Once it is established that the coarsening stage of polymer collapse is indeed a scaling phenomenon, the next interest goes 
towards checking the associated growth laws. In Fig.\ \ref{cluster_size}(a), we show the time dependence of $C_s(t)$ for OLM and LM. 
To make the data from both models visible on the same scale there the $y$-axis is scaled by the corresponding chain length $N$ of the polymer. 
Note that saturation of the data for LM at a value less than unity is due to the fact that there we have calculated the average cluster size $C_s(t)$ 
from the decay of the correlation function $C(r,t)$ as described in the previous subsection. This gives a proportionate measure of the average 
number of monomers present in the clusters and thus the data saturate to a value less than unity.
\par
In coarsening kinetics 
of binary mixtures such time dependence of the relevant length scale can be described correctly when one 
considers an off-set in the scaling ansatz \cite{Majumder2011,Majumder2010,das2012,majumder2013}. Similarly, it was later proved to 
be appropriate for the cluster growth during the collapse of 
a polymer \cite{MajumderEPL,majumder2017SM}. Following this one writes down the scaling ansatz as 
\begin{eqnarray}\label{cl_growth}
C_s(t)=C_0+At^{\alpha_{c}},
\end{eqnarray}
where $C_0$ corresponds to the cluster size after crossing over from the initial cluster formation stage, and 
$A$ is a temperature-dependent amplitude. The solid lines in Fig.\ \ref{cluster_size}(a) are fits to the form \eqref{cl_growth} 
yielding $\alpha_c =0.98(4)$ and $0.62(5)$ for OLM and LM, respectively. 
\par
One can verify the robustness of the growth by studying the dependence of cluster growth on the quench temperature $T_q$. 
For this one uses data at different $T_q$ and can perform a scaling analysis based on 
nonequilibrium finite-size scaling (FSS) arguments \cite{majumder2017SM}. The nonequilibrium FSS analysis was constructed based 
on FSS analyses in the context of equilibrium critical phenomena \cite{Fisherbook,Privmanbook}. An account of the FSS formulation in the present context 
can be found in Ref.\ \cite{majumder2017SM}. In brief, one introduces in the growth ansatz \eqref{cl_growth} a 
scaling function $Y(y_p)$ as 
\begin{eqnarray}\label{FS_ansatz}
C_s(t)-C_0=(C_{\max}-C_0)Y(y_p),
\end{eqnarray}
which implies
\begin{eqnarray}\label{FS_func_cl}
Y(y_p)=\frac{(C_s(t)-C_0)}{(C_{\max}-C_0)},
\end{eqnarray}
where $C_{\max} \sim N$ is the maximum cluster size a finite system can attain. In order to account for the 
temperature-dependent amplitude $A(T_q)$, one uses the scaling variable 

\begin{eqnarray}\label{FS_variable_T}
y_p= f_s\frac{(N-C_0)^{1/\alpha_{c}}}{(t-t_0)}
\end{eqnarray}
where
\begin{eqnarray}\label{FS_variable_T2}
f_s=\left[\frac{A(T_{q,0})}{A(T_q)}\right]^{1/\alpha_c}.
\end{eqnarray}
The metric factor $f_s$ is introduced for adjusting the nonuniversal amplitudes $A(T_q)$ at different $T_q$. Here, in addition to $C_0$ one also 
uses the crossover time $t_0$ from the initial cluster formation stage. 
A discussion of the estimation of $C_0$ and $t_0$ can be found in Refs.\ \cite{majumder2017SM,christiansen2017JCP}. While performing the exercise 
we tune the parameters $\alpha_c$ and $f_s$ to obtain a data collapse along with the $Y(y_p) \sim y_p^{-\alpha_c}$ behavior 
in the finite-size unaffected region. In Figs.\ \ref{cluster_size}(b) and (c), we demonstrate such scaling exercises with $\alpha_c=1.0$ and $0.62$ for OLM and LM, respectively. For $f_s$, we use the reference temperature $T_{q,0}=1.0$ and $2.0$ for OLM and LM, respectively. The collapse of data for different $T_q$ and consistency with the corresponding $y_p^{-\alpha_c}$ behavior in both plots 
suggest that the growth is indeed quite robust and can be described by a single finite-size scaling function with nonuniversal metric factor $f_s$ 
in the scaling variable. However, $\alpha_c$ in OLM is larger than for LM, a fact in concurrence with the values 
of $z$ estimated previously, and thus to some extent providing a support to the heuristic relation $z \sim 1/\alpha_c$. The use of a nonuniversal metric factor in order 
to find out an universal FSS function was first introduced in the context of equilibrium critical phenomena using different lattice types \cite{Privman1984,hu1995}.
After adapting this concept to nonequilibrium FSS of polymer kinetics in Refs.\ \cite{majumder2017SM,christiansen2017JCP} as explained above, it was recently also transferred
to spin systems where its usefulness has been demonstrated in a coarsening study of the Potts model with conserved dynamics \cite{majumder2018_potts}.

\subsection{Aging and related scaling}\label{aging}
Apart from the scaling of the growing length scale or the cluster size that deals only with 
equal-time quantities, coarsening processes are associated with the aging phenomenon as well. Thus along the same line, in order to 
check aging during collapse of a polymer one can calculate the two-time correlation function or the autocorrelation function described in 
Eq.\ \eqref{auto_cor}. However, unlike for spin systems here the choice of the observable $O_i$ is not trivial. 
Nevertheless, for OLM we identified the observable $O_i$ as a variable 
based on the cluster identification method. We assign $O_i = \pm1$ depending on whether 
the monomer is inside ($+1$) or outside ($-1$) a cluster. It is apparent that our cluster 
identification method is based on the local density around a monomer along the chain. 
Thus $C(t,t_w)$ calculated using this framework gives an analogue of the usual 
density-density autocorrelation functions in particle systems. On the other hand for LM,
we assign $O_i=\pm 1$ by checking the radius $r$ at which the local density, given by $\rho_i(r,t)$ 
[see Eqs.\ \eqref{cor_lattice} and \eqref{rho_lattice}], first falls below a threshold of $0.1$. 
If this radius is smaller than $\sqrt{3}$ we assign $O_i=1$, marking a high local density, otherwise we 
chose $O_i=-1$ to mark a low local density. For details see Refs.\ \cite{majumder2017SM} and \cite{christiansen2017JCP} for OLM and LM, respectively. 
\begin{figure}[t!]
\centering
\resizebox{0.9\columnwidth}{!}{\includegraphics{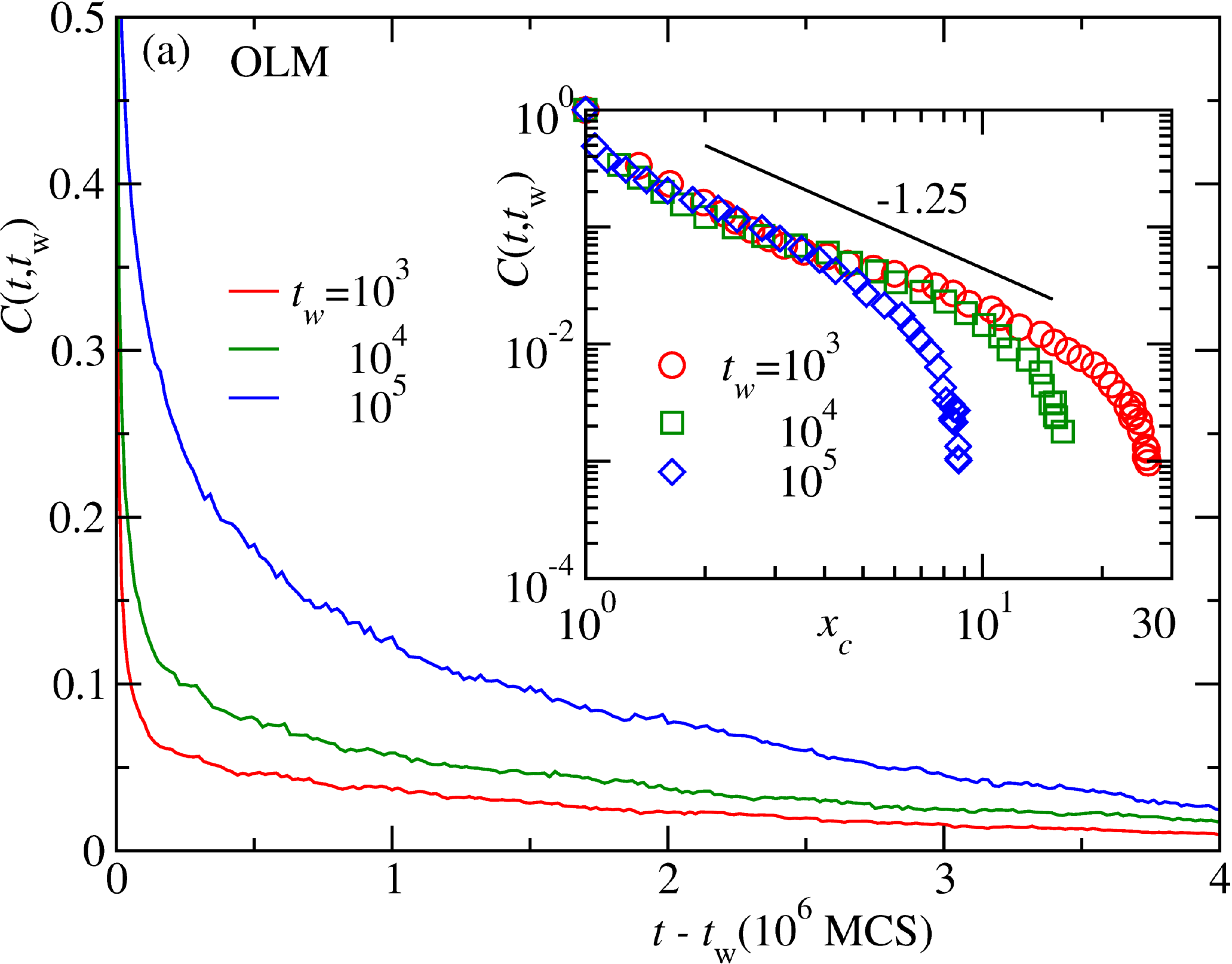}}
\vskip 0.1cm
\resizebox{0.9\columnwidth}{!}{\includegraphics{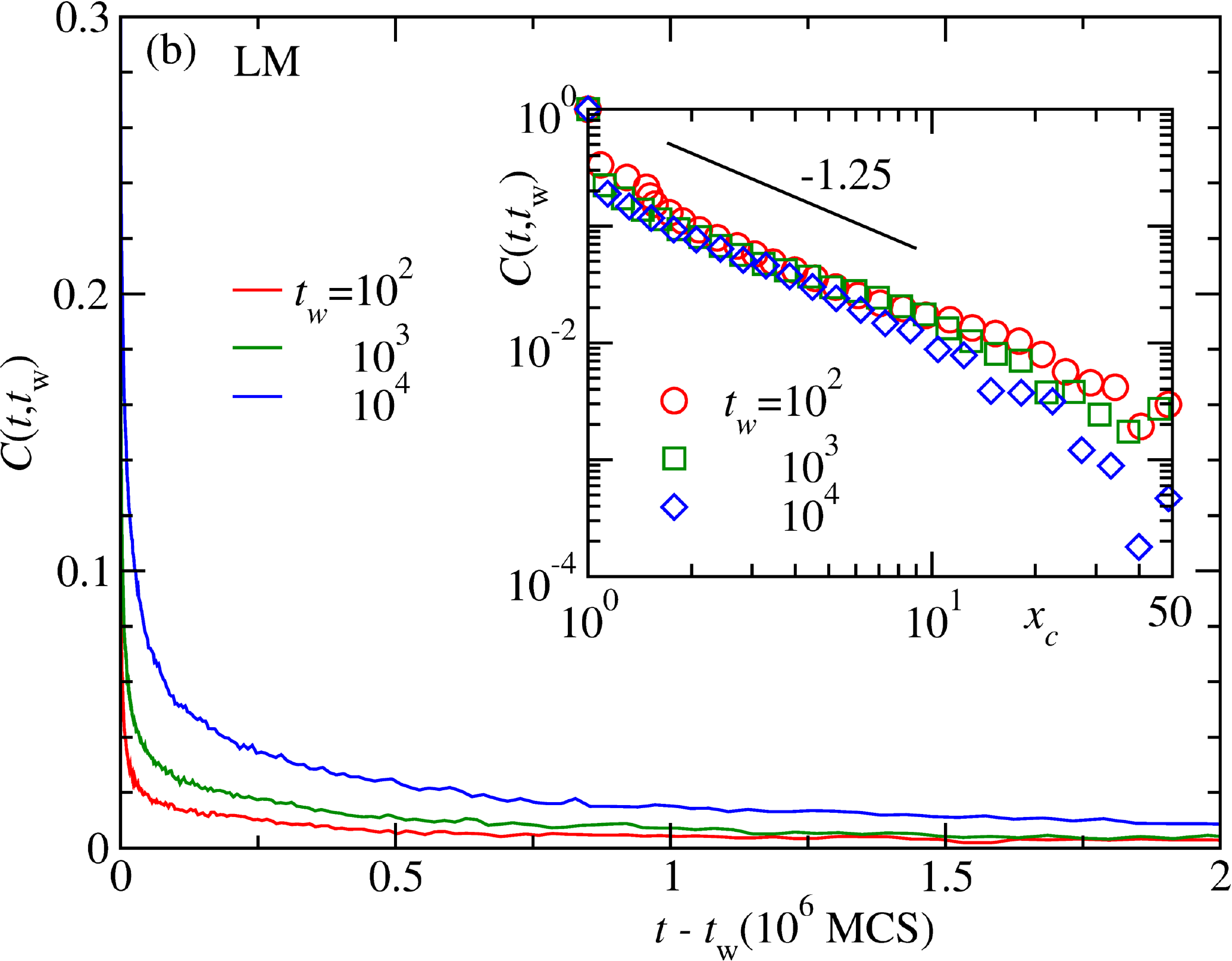}}
\caption{ Demonstration of aging phenomenon during collapse of a polymer for (a) OLM and (b) LM. The main frame shows the 
plot of the autocorrelation functions calculated using \eqref{auto_cor} at different waiting times $t_w$, as mentioned there. 
The insets show the corresponding scaling plots with respect to the scaling variable $x_c=C_s(t)/C_s(t_w)$, in accordance with \eqref{power-law_Cst}.  
The solid lines depict the consistency of the data with a power law having an exponent $\lambda_c=1.25$.}
\label{corr_lw}      
\end{figure}
\par
In the main frames of Figs.\ \ref{corr_lw}(a) and (b) we show plots of the autocorrelation function $C(t,t_w)$ against the translated 
time $t-t_w$ for (a) OLM and (b) LM. Data from both the cases clearly show breaking of time-translation invariance, one of the necessary 
conditions for aging. It is also evident that as $t_w$ increases, the curves decay more slowly, an indication of slow relaxation behavior fulfilling the 
second necessary condition for aging. For the check of the final condition for aging, i.e., dynamical scaling, in principle one could study
the scaling with respect to the scaled time $t/t_w$. Although such an exercise provides a reasonable collapse of data 
for OLM, data for LM do not show scaling with respect to $t/t_w$. In this regard, one could look for special aging 
behavior that can be achieved by considering \cite{henkelbook}
\begin{equation}\label{superaging}
 C(t,t_w) \equiv G\left(\frac{h(t)}{h(t_w)}\right),
\end{equation} 
with the scaling variable
\begin{equation}
h(t)=\exp\left(\frac{t^{1-\mu}-1}{1-\mu}\right).
\end{equation}
Here, $G$ is the scaling function and $\mu$ is a nontrivial exponent.
Special aging with $0 <\mu < 1$ is referred to as subaging and has been observed mostly in soft-matter systems \cite{Cloitre2000,bursac2005,wang2006}, in spin glasses \cite{hilhorst1981,herisson2004,Parker2006}, and recently 
in long-range interacting systems \cite{christiansen2019non}. The $\mu>1$ case is referred to as superaging and was claimed to be observed in site-diluted Ising ferromagnets. 
However, Kurchan's lemma \cite{Kurchan2002} rules out the presence of apparent superaging \cite{paul2007}. This was further consolidated via numerical 
evidence in Ref.\ \cite{Park2010}. There it has been 
argued that the true scaling is observed in terms of the ratio of growing length scales at the corresponding times, i.e., $\ell(t)/\ell(t_w)$. 
In the case of polymer collapse with LM, too, one apparently observes special scaling of the form \eqref{superaging} with $\mu < 1$, i.e., subaging in this case. However, 
following the argument of Park and Pleimling \cite{Park2010}, one gets also here the simple scaling behavior with respect to the scaling variable $x_c=C_s(t)/C_s(t_w)$, thus ruling out the presence 
of subaging. Such scaling plots of the autocorrelation data both for OLM and LM are shown in the insets of Fig.\ \ref{corr_lw}. In both 
cases the data seem to follow the power-law scaling with a decay exponent $\lambda_c \approx 1.25$. 
\begin{figure}[t!]
\centering
\resizebox{0.9\columnwidth}{!}{\includegraphics{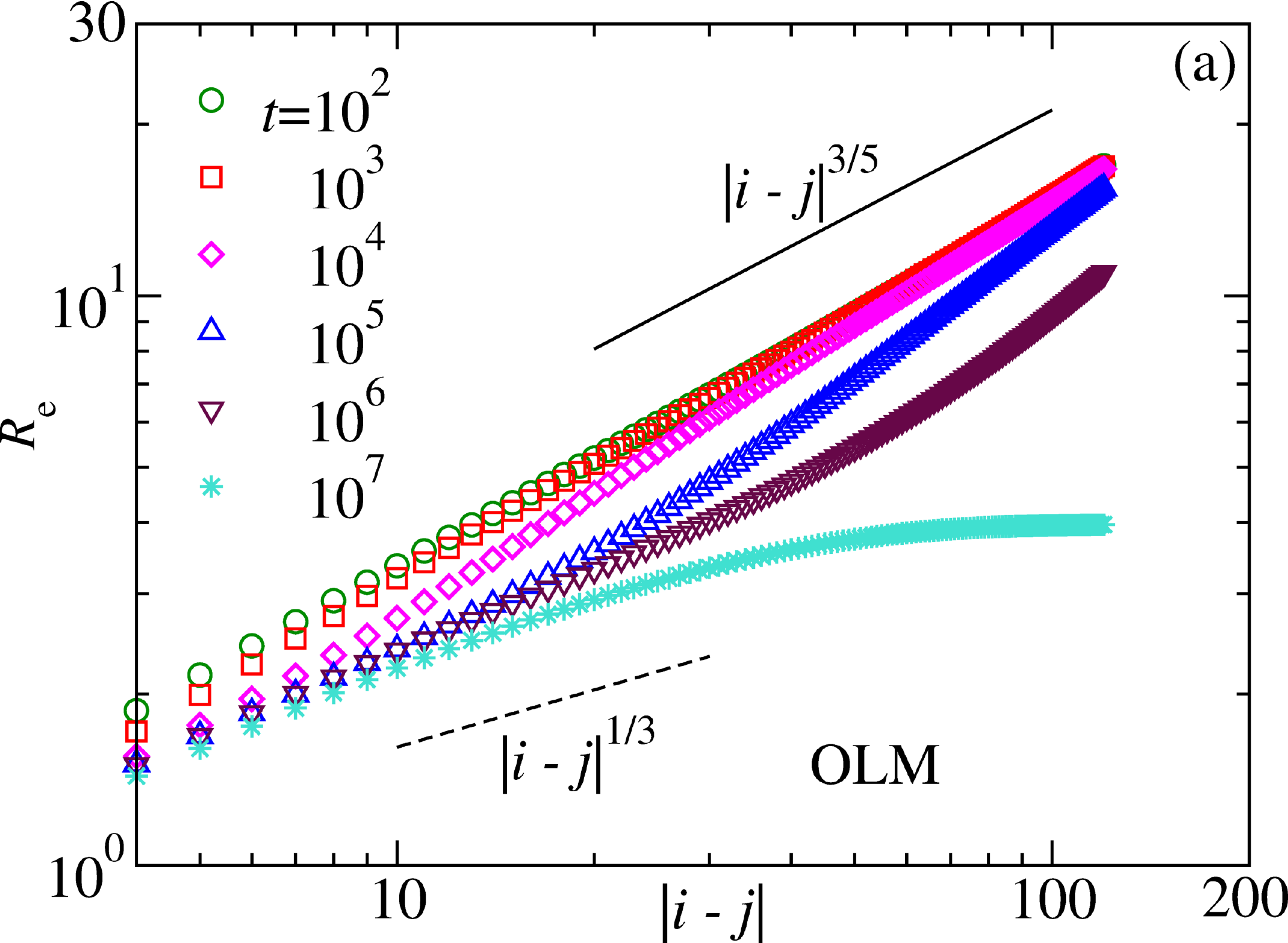}}
\vskip 0.1cm
\resizebox{0.9\columnwidth}{!}{\includegraphics{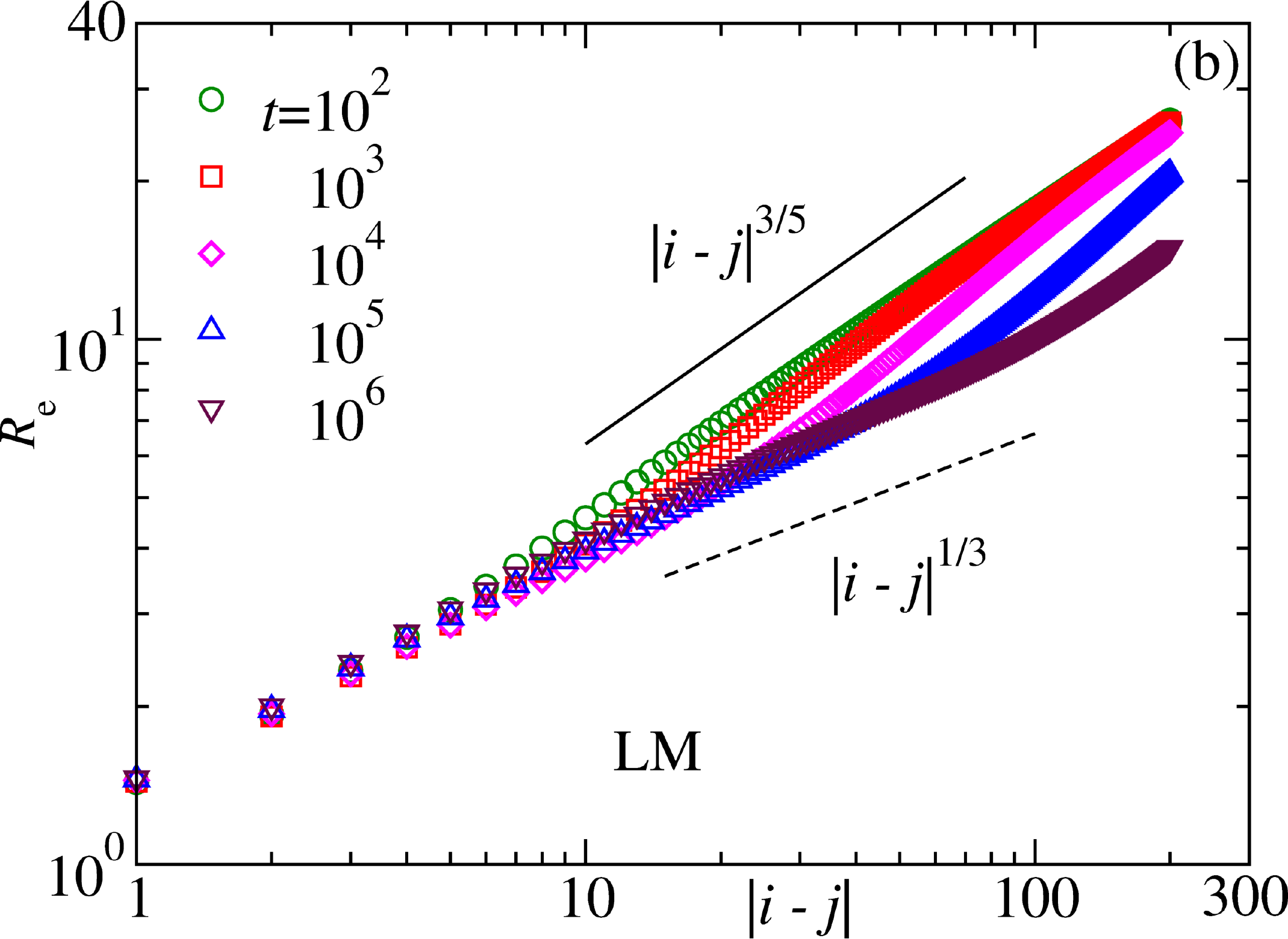}}
\caption{ Geometrical distance between monomers $i$ and $j$ which are at a distance $|i-j|$ 
along the contour of the chain for a polymer using (a) OLM and (b) LM, at different times mentioned. The respective chain lengths are $N=724$ and $2048$ and the 
quench temperatures are $T_q=1.0$ and $2.5$.  The solid line shows the expected behavior for an extended coil and the dashed line 
shows the behavior in the collapsed phase. The plot in (a) is taken from Ref.\ \cite{majumder2017SM}.}
\label{rg_vs_Cs}      
\end{figure}
\par
Relying on the fact that the calculation of $C(t,t_w)$ is based on the cluster identification criterion, 
i.e., by calculating the local monomer densities around each monomer along the polymer chain, it gives 
an analogue to the usual density-density autocorrelation function as used in glassy systems. 
Keeping in mind the corresponding argument for the bounds on the respective aging exponent 
for spin-glass and ferromagnetic ordering, one can thus assume \cite{Majumder2016PRE} $C(t,t_w) \sim \langle \rho(t) \rho(t_w) \rangle$
where $\rho$ is the average local density of monomers. Now let us consider a set of $C_s$ monomers 
at $t~(\gg t_w)$ and assume that at $t_w$ the polymer is more or less in an extended coil 
state where the squared radius of gyration scales as $R_{g}^2 \sim N^{2\nu_F}$. Using $C_s \equiv N$ 
in this case one can write 
\begin{eqnarray}\label{den_tw}
\rho(t_w) \sim C_s/{R_g}^{d} \sim C_s^{-(\nu_F d-1)}.
\end{eqnarray}
The above fact can be verified from Figs.\ \ref{rg_vs_Cs}(a) and (b) for OLM and LM, respectively, where we plot the average geometrical (Euclidean) 
distance $R_e$ ($\sim R_g$) between the monomers $i$ and $j$ placed at a distance $|i-j|$ along the contour 
of the chain at different times during the collapse.
For both cases, the data at early times show that the behavior is consistent with an extended coil governed by the Flory exponent $\nu_F=3/5$. This consolidates the foundation of the relation \eqref{den_tw} 
provided $t_w$ is at early times. 
\begin{figure}[t!]
\centering
\resizebox{0.95\columnwidth}{!}{\includegraphics{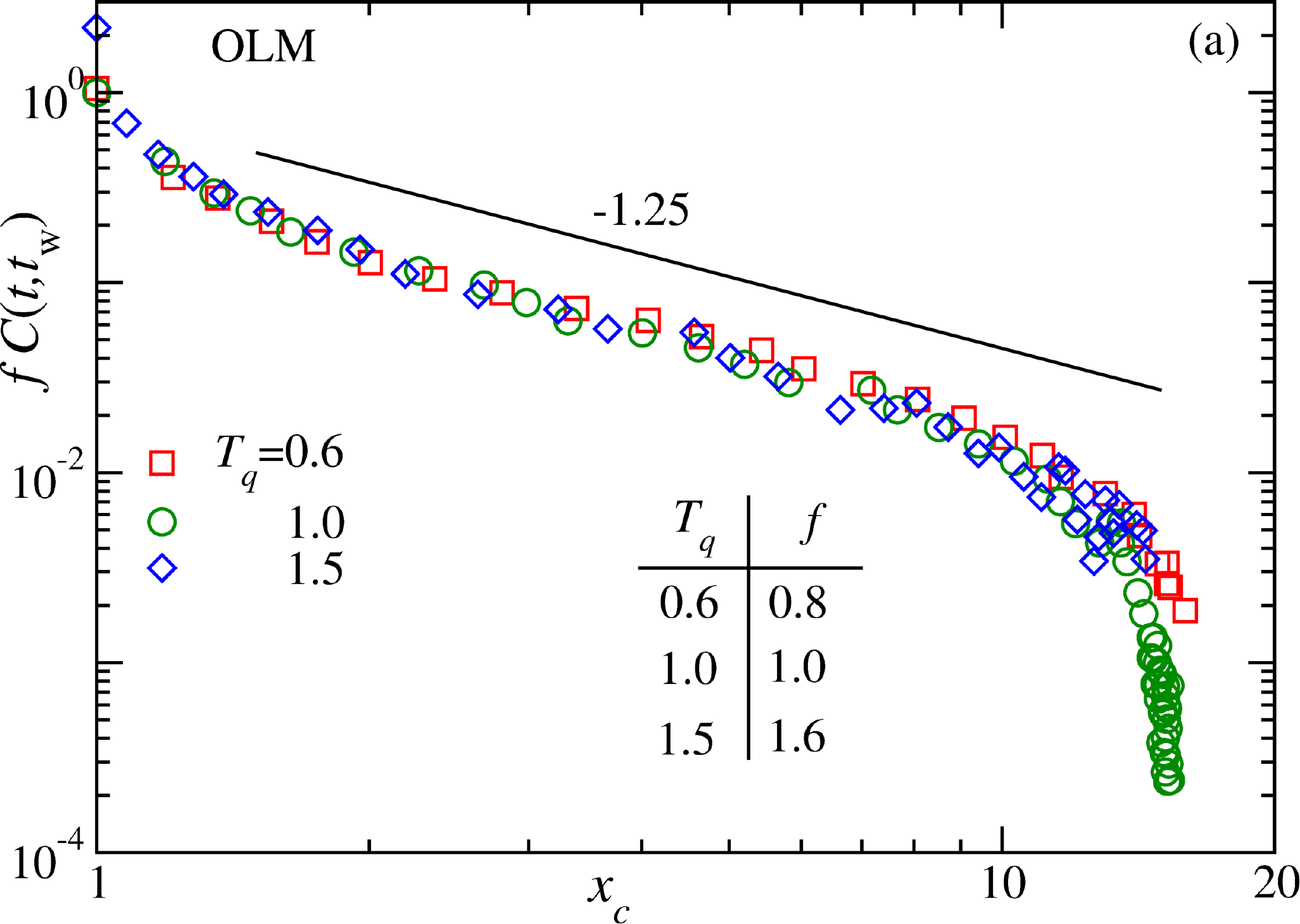}}
\vskip 0.1cm
\resizebox{0.95\columnwidth}{!}{\includegraphics{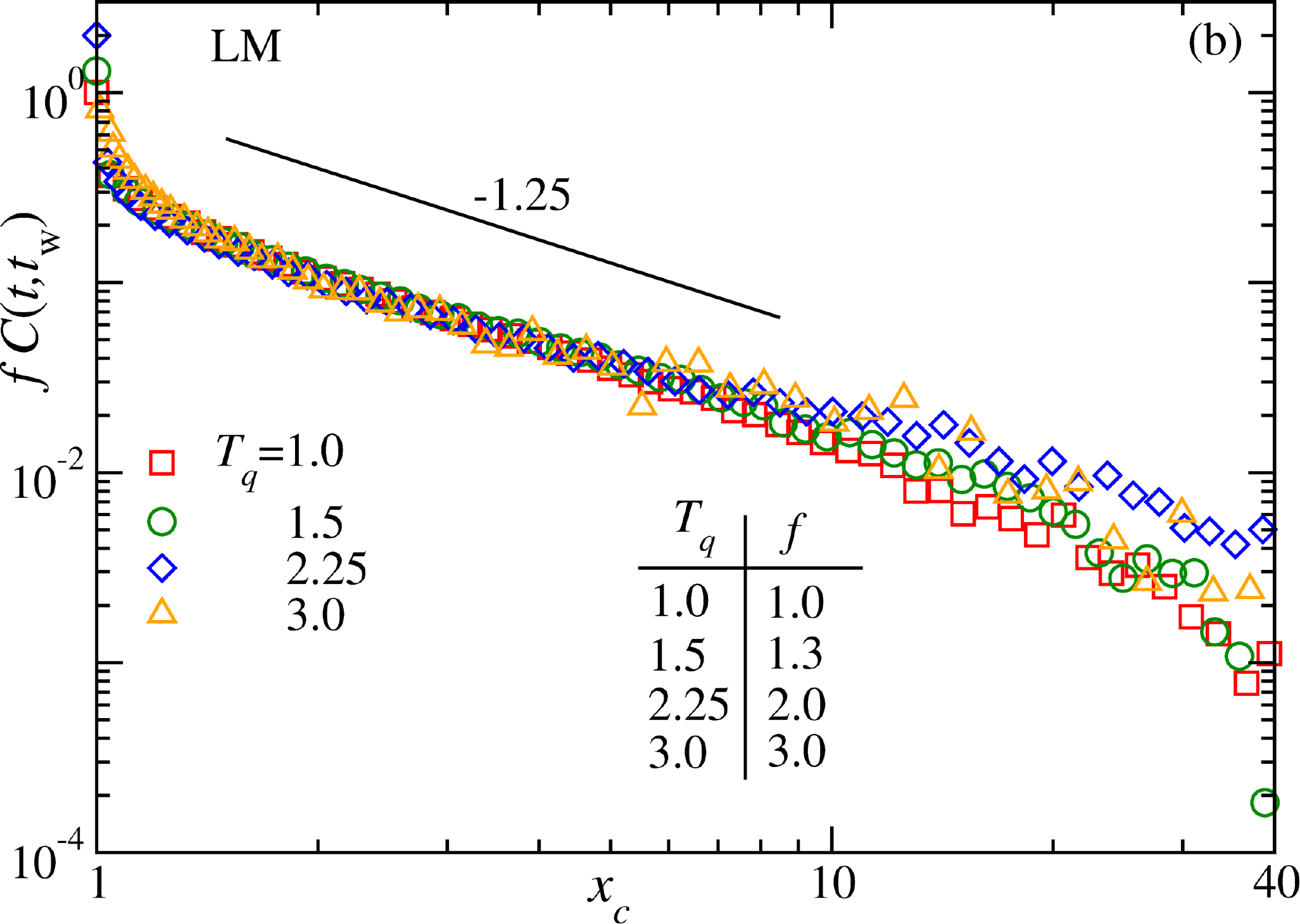}}
\vskip 0.1cm
\resizebox{0.95\columnwidth}{!}{\includegraphics{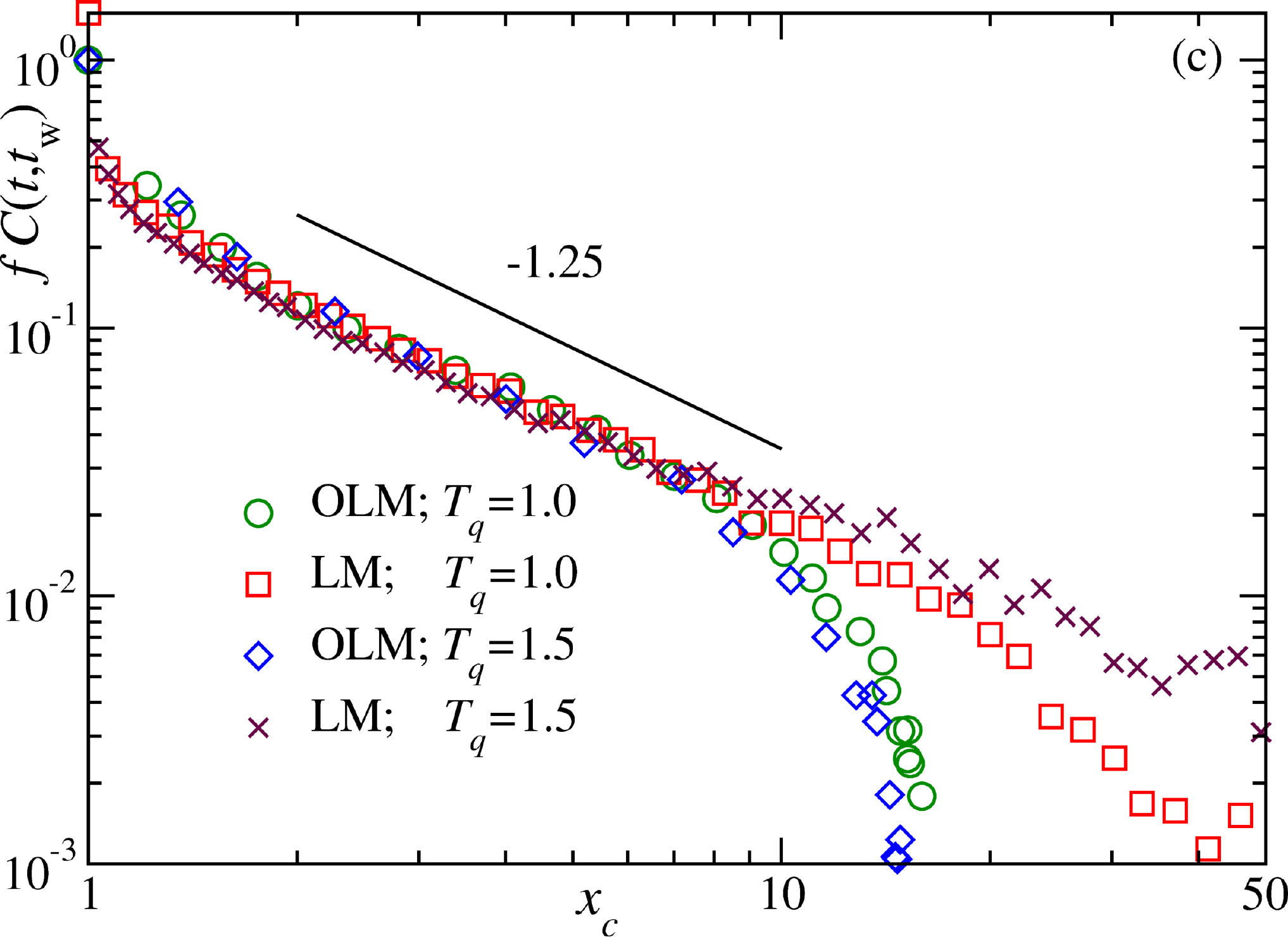}}
\caption{ Plots demonstrating that aging scaling of the autocorrelation function $C(t,t_w)$ at different $T_q$ for (a) OLM and (b) LM can be described 
by a single master curve when plotted as a function of $x_c=C_s(t)/C_s(t_w)$. The solid lines there again correspond to \eqref{power-law_Cst} with $\lambda_c=1.25$. 
For OLM, the used data are at $t_w=5\times 10^3$, $10^4$ and $3\times 10^4$, respectively, for $T_q=0.6$, $1.0$ and $1.5$. 
For LM, data for all temperatures are at $t_w \approx 10^3$. Note that here we have simply multiplied the $y$-axis by a factor $f$ to make the 
data fall onto the same master curve. (c) Illustration of the universal nature of aging scaling in the two models. Here the used data  
are at $t_w=10^4$ and $10^3$ for OLM and LM, respectively. Adapted from Refs.\ \cite{majumder2017SM,christiansen2017JCP,majumder2018proceeding}.}
\label{corr_lw_comp}      
\end{figure}
\par
Now at the observation time $t$ there are two possibilities. 
Firstly, if $t$ is late enough, then we expect that all the monomers will be inside a 
cluster which gives $R_g\sim C_s^{1/d}$ so that $\rho(t)=1$. Thus considering the maximum 
overlap between $\rho(t)$ and $\rho(t_w)$ we get 
\begin{eqnarray}\label{lower_bound}
 C(t,t_w) \sim C_s^{-(\nu_F d-1)}. 
\end{eqnarray}
This gives the lower bound. Secondly, with the assumption that the polymer is in an extended coil 
state even at time $t$, then $\rho(t)=\rho(t_w)$ holds and we obtain
\begin{eqnarray}\label{upper_bound}
C(t,t_w) \sim C_s^{-2(\nu_F d-1)},
\end{eqnarray}
providing the upper bound for the aging exponent $\lambda_c$. Thus by combining 
\eqref{lower_bound}  and \eqref{upper_bound} we arrive at the bounds \cite{Majumder2016PRE}
\begin{eqnarray}\label{poly-bound}
 (\nu_F d-1) \le \lambda_c \le 2(\nu_F d-1).
\end{eqnarray}
Putting $\nu_F=3/5$ in \eqref{poly-bound} one would get $4/5 \le \lambda_c \le 8/5 $. 
Further, inserting the more precise numerical estimate in $d=3$ as \cite{clisby2010,Clisby2016}
$\nu_F=0.587\,597$, we get 
\begin{eqnarray}\label{The-bound}
 0.762\,791\le \lambda_c \le 1.525\,582.
 \end{eqnarray}
The validity of this bound can also be readily verified from the consistency 
of our data in the insets of Fig.\ \ref{corr_lw} with the solid lines having a power-law decay with 
exponent $1.25$.  We make the choice of $t_w$ in all the plots so that the assumption 
that at time $t_w$ the polymer is in an extended coil state is valid. This choice can also be 
appreciated from the plots in Figs.\ \ref{rg_vs_Cs}(a) and (b) for OLM and LM, respectively. There it is evident that the extended coil behavior ($R_e \sim |i-j|^{3/5}$) at early times is gradually changing to the 
behavior expected for the collapsed phase ($R_e \sim |i-j|^{1/d}$ with $d=3$) at late times.  
The little off behavior of the data for higher $t_w$ in the inset of Fig.\ \ref{corr_lw} is indeed due to the 
fact that at those times the formation of stable clusters has already initiated to change the extended
coil behavior of the chain. Confirmation of the value of $\lambda_c$ via finite-size scaling can also be done as 
presented in Refs.\ \cite{Majumder2016PRE,christiansen2017JCP}.
\par
To confirm the robustness of the above bound and the value of $\lambda_c$, we plot $C(t,t_w)$ from different temperatures $T_q$ in Fig.\ \ref{corr_lw_comp}(a) for OLM and  Fig.\ \ref{corr_lw_comp}(b) for LM. Mere plotting of those data yields curves that are parallel to each other 
due to different amplitudes. However, if one uses a multiplier $f$ on the $y$-axis to adjust those  
different amplitudes for different $T_q$ one obtains curves that fall on top of each other as shown.
The values of $f$ used for different $T_q$ are mentioned in the tables within the plots. Note that this non-trivial factor $f$ is similar to the 
nonuniversal metric factor $f_s$ used for the cluster growth in the previous subsection. The solid lines in both the cases show the consistency 
of the data with the scaling form \eqref{power-law_Cst} with $\lambda_c=1.25$. To further check the universality of the 
exponent $\lambda_c$ we now compare the results from aging scaling obtained for the polymer collapse using the two polymer models.
For that we plot in Fig.\ \ref{corr_lw_comp}(c) the data for different $T_q$ coming from both models on the same graph. Here again, we have used
the multiplier $f$ for the data collapse. Collapse of data irrespective of the model and the temperatures $T_q$ onto a master-curve behavior 
and their consistency with the power-law scaling \eqref{power-law_Cst} having $\lambda_c=1.25$ (shown by the solid line), speaks for 
the universal nature of aging scaling during collapse of a polymer. 
\begin{figure*}[t!]
\centering
\resizebox{0.8\textwidth}{!}{\includegraphics{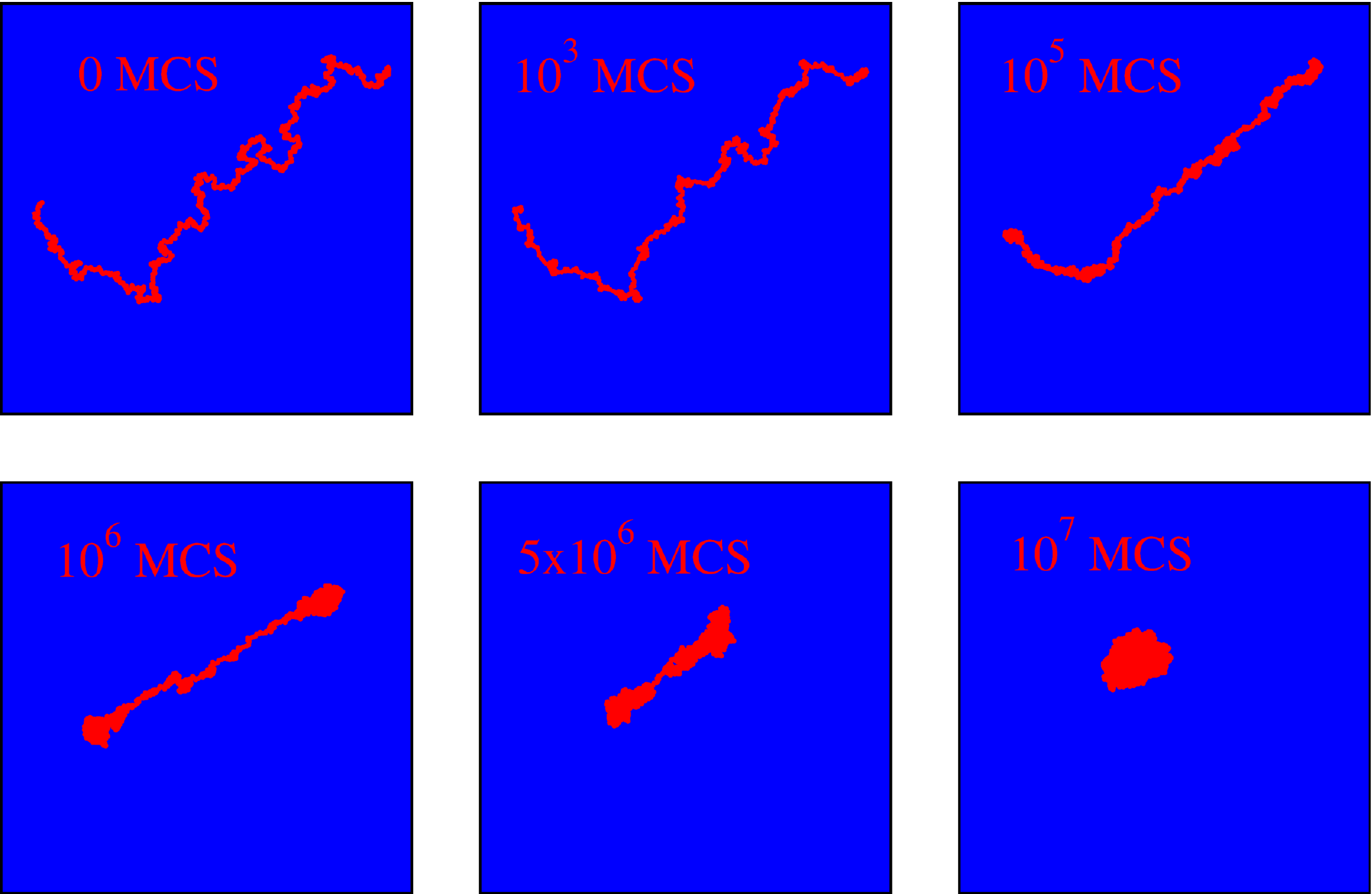}}
\caption{ Plot showing the time evolution of a polymer in $d=2$ using OLM after being quenched from a high-temperature extended coil phase to 
a temperature $T_q=1.0$ where the equilibrium phase is globular. The times are mentioned in there and the used chain length is $N=512$.}
\label{snap2d}      
\end{figure*}
\section{Results for the case of OLM in $d=2$ }\label{2D}
In this section we present some preliminary results for the kinetics of polymer collapse in $d=2$ dimensions using only OLM as defined 
by Eqs.\ \eqref{FENE}, \eqref{potential_OLM}, and \eqref{std_LJ}. Experiments on polymer dynamics are often set up by using an 
attractive surface which effectively confines the polymer to move 
in two-dimensional space. Thus understanding the scenario in pure $d=2$ dimensions provides some impression about 
such quasi-two-dimensional geometry \cite{deGennesbook,vanderzandebook}. From a technical point of view, simple Metropolis simulations of a polymer in $d=2$ are 
much more time consuming than in $d=3$. This is due to the absence of one degree of freedom which makes the collapse of the polymer 
difficult via local moves and thereby increasing the intrinsic time scale of collapse. In fact even in equilibrium there are very few 
studies \cite{wittkop1996,polson2000,Grassberger2002,Zhou2006} and in particular we do not find any study that gives an idea about the 
collapse transition temperature. Since for the study of the 
kinetics the actual value of the transition temperature is not crucial we performed a few equilibrium simulations in $d=2$ covering a wide range of temperatures and 
found that at $T_q=1.0$ the polymer is in the collapsed phase for a chain length of $N=512$, while it remains in an extended coil state at $T_h=10.0$. 
So for this work we have used a polymer of length $N=512$ and prepared an initial configuration at $T_h=10.0$ before quenching it to a temperature 
$T_q=1.0$. All the other specifications for the simulation method remain the same as we discussed it for OLM in Section\ \ref{model}, apart from
confining the displacement moves to only $d=2$ dimensions. 

\par
In Fig.\ \ref{snap2d} we show the time evolutions during the collapse of the $d=2$ polymer at $T_q=1.0$. The sequence of events 
portrayed by the snapshots shows formation of local ordering as observed for $d=3$, although the formation of a ``pearl-necklace'' is 
not so evident. By comparing with the snapshots presented for $d=3$ in Figs.\ \ref{figsnap} and \ref{figschematic}, it is apparent 
that the initial process of local cluster formation 
is much slower in $d=2$. However, once the local clusters are formed (as shown in the snapshot at $t=10^6$ MCS) the time evolution shows coarsening of these 
clusters to finally form a single cluster or globule. Thus the overall phenomenology seems to be in line with the $d=3$ case.
\begin{figure}[b!]
\centering
\resizebox{0.95\columnwidth}{!}{\includegraphics{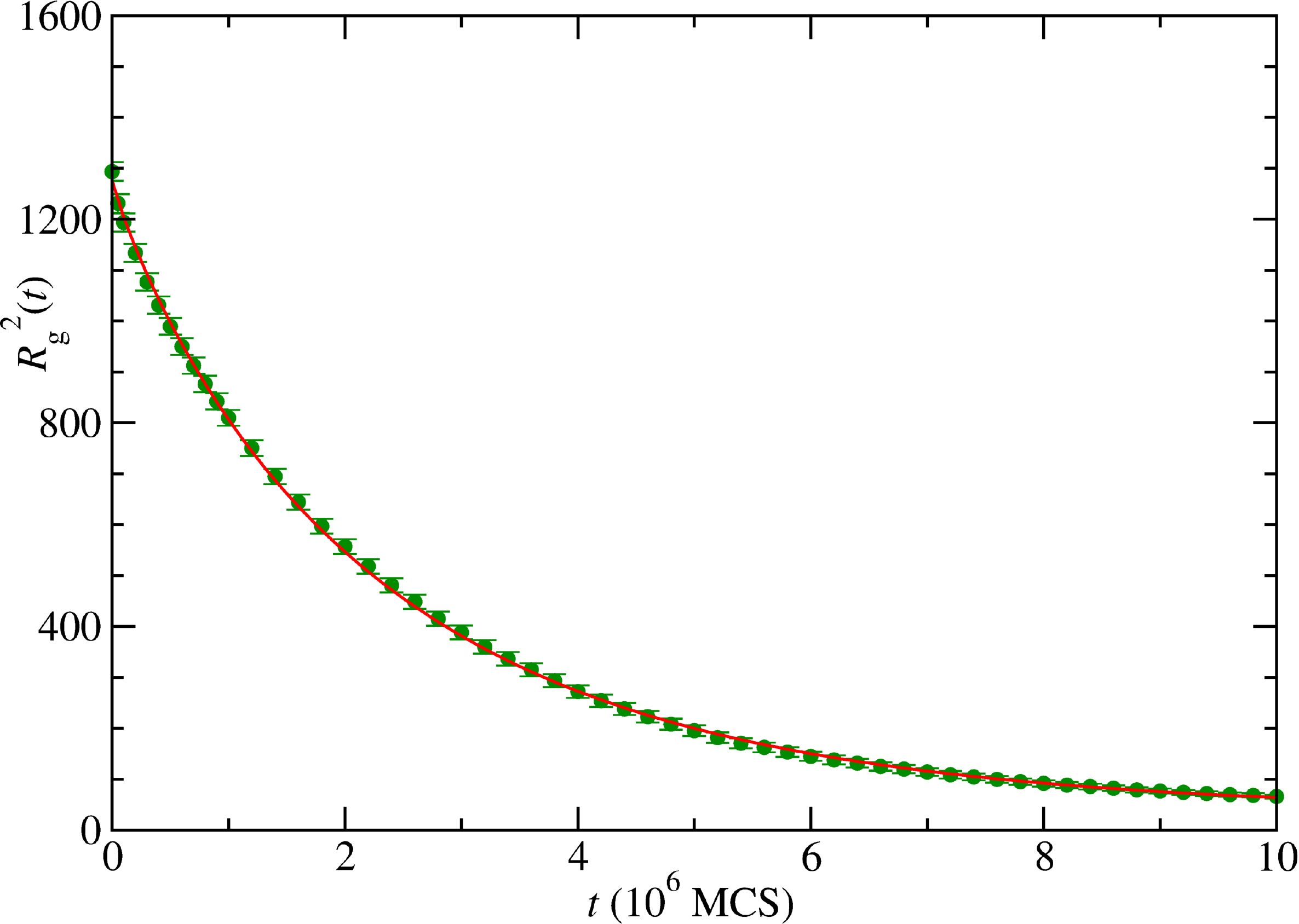}}
\caption{ Time dependence of the average squared radius of gyration $R_g^2$ during collapse of a polymer in $d=2$. The system size and the quench temperature 
are the same as in Fig.\ \ref{snap2d}. The continuous line is a fit to the data using Eq.\ \eqref{rg_decay}.}
\label{rg2d}      
\end{figure}
\par
Following what has been done for the $d=3$ case, at first we look at the time dependence of the overall size of the polymer 
by monitoring the squared radius of gyration $R_g^2$. In Fig.\ \ref{rg2d} we show the corresponding plot of $R_g^2$ (calculated as an 
average over $300$ different initial realizations). Like in the $d=3$ 
case, the decay of $R_g^2$ can be described quite well via the empirical relation mentioned in Eq.\ \eqref{rg_decay}. The best fit 
obtained is plotted as a continuous line in the plot. The obtained value of the non-trivial parameter $\beta$ in this fitting is $\approx 0.89$, which is compatible with the $d=3$ case \cite{majumder2017SM}. Still, the dependence of $\beta$ on the chain length $N$ would be worth investigating and will 
be presented elsewhere. Along the same line an understanding of the scaling of the collapse time with the chain length will be interesting to 
compare with the $d=3$ case. As this Colloquium is focused more on the cluster coarsening and aging during the collapse, here, we abstain ourselves from 
presenting results concerning the scaling of the collapse time. 

\subsection{Cluster coarsening }
\begin{figure}[t!]
\centering
\resizebox{0.9\columnwidth}{!}{\includegraphics{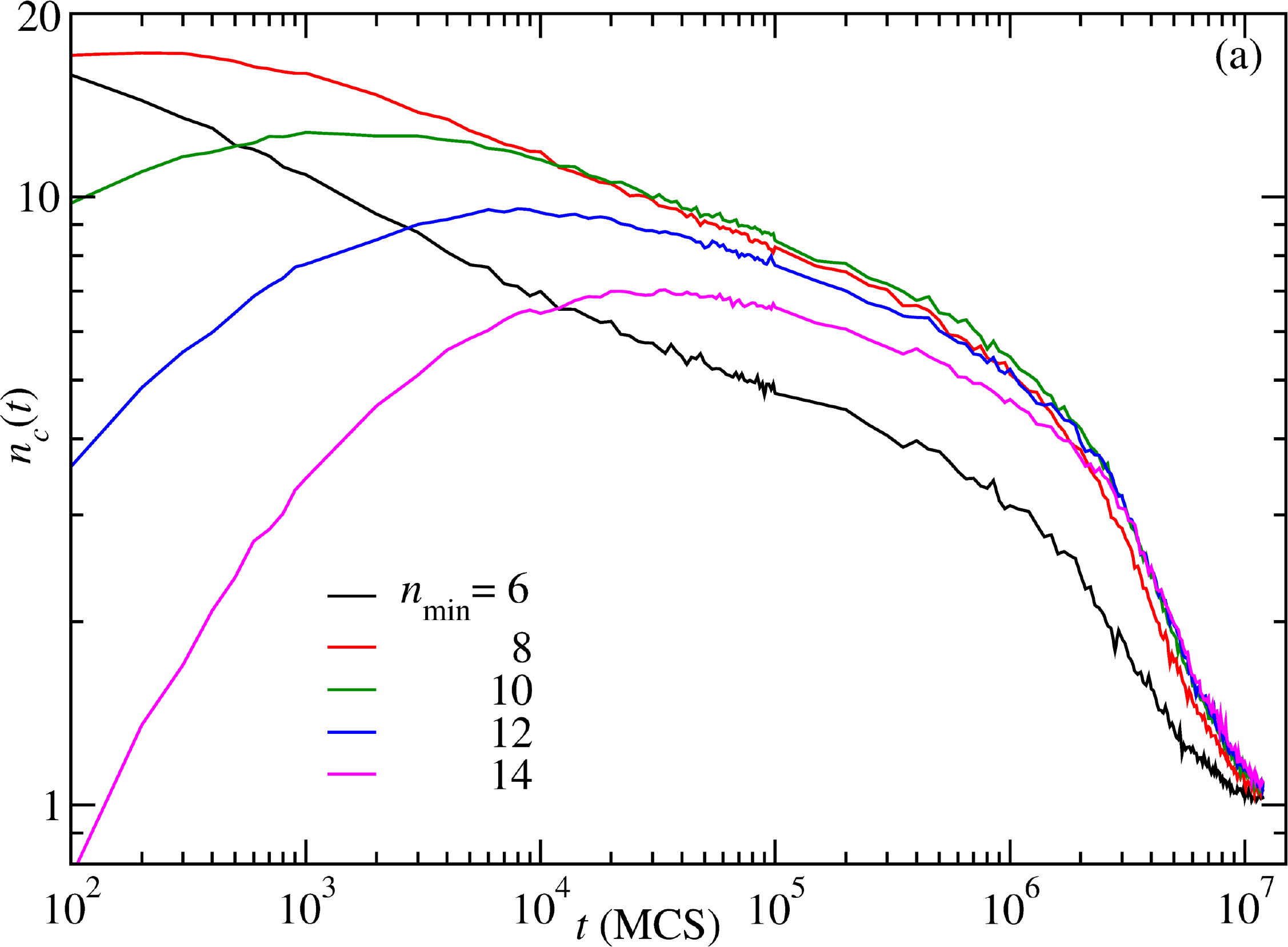}}
\vskip 0.1 cm
\resizebox{0.9\columnwidth}{!}{\includegraphics{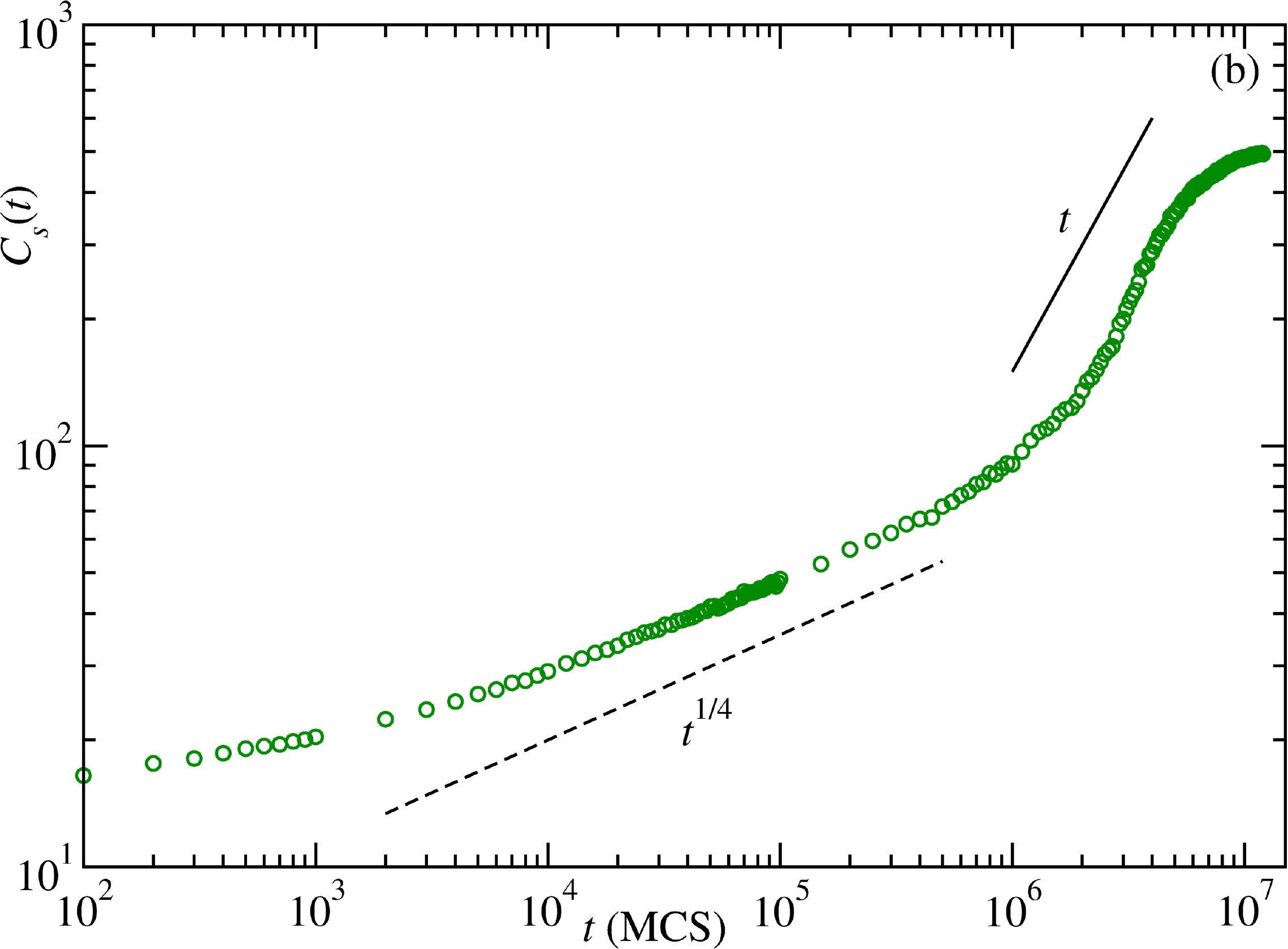}}
\caption{ (a) Plot of the average number of clusters of monomers $n_c$ as a function of time during collapse of a polymer with chain length $N=512$ 
modeled via OLM in $d=2$ at $T_q=1.0$. Results for different choices of $n_{\min}$ are shown demonstrating the late-time consistency of the 
data with each other. (b) Illustration of the scaling of the cluster growth during collapse via a plot of the average cluster size $C_s$ as 
a function of time. Here we have used $n_{\min}=12$. The dashed and the solid lines correspond to different power-law behaviors observed 
at early and late times, respectively.}
\label{noc2d}      
\end{figure}

As can be seen from the snapshots in Fig.\ \ref{snap2d}, during the course of the collapse, like in $d=3$, also for $d=2$ one notices formation of 
local clusters which via coalescence with each other form bigger clusters and eventually form a single cluster or globule. We measure the 
average cluster size in the following way. First we calculate the total 
numbers of monomers in the nearest vicinity of the $i$-th monomer as 
\begin{eqnarray}\label{cl_identify}
n_i=\sum\limits_{j=1}^{N}\Theta (r_{c}-r_{ij}), 
\end{eqnarray}
where $r_c$ is the same cutoff distance used in the potential \eqref{potential_OLM} for the simulations  and $\Theta$ is the 
Heaviside step function. For $n_i \ge n_{\rm{min}}$, 
there is a cluster around the $i$-th monomer and all those $n_i$ monomers belong to that 
cluster. The total number of clusters calculated this way may include some overcounting, 
which we remove via the corresponding Venn diagram, and thus the actual discrete 
clusters $k=1,\dots, n_c(t)$ are identified and the number of monomers $m_{k}$ within each cluster 
is determined. Finally the average cluster size is calculated as 
\begin{eqnarray}
C_s(t) = \frac{1}{n_c(t)} \sum \limits_{k=1}^{n_c(t)} m_k,
\end{eqnarray}
where $n_c(t)$ is the total number of discrete clusters at time $t$. Note that in this calculation 
we do not vary the cut-off radius $r_c$ and fix it to the same value ($r_c=2.5\sigma$) as we have used for our simulations. Hence, the obtained value of $C_s(t)$ depends only on one nontrivial choice, which is $n_{\rm{min}}$. Figure\ \ref{noc2d}(a) shows 
how the identification of clusters depends on different choices of $n_{\min}$ during collapse 
of a polymer having length $N=512$. There we have plotted the 
average number of clusters as a function of time for different $n_{\min}$. One can notice for choices of 
$n_{\min} \ge 10$ the late-time behaviors are more or less indistinguishable. However, the initial 
structure formation stage is well covered by the choice $n_{\min}=12$. Thus we consider $n_{\min}=12$ 
as the optimal value to identify and calculate the average cluster size. 
\par
In Fig.\ \ref{noc2d}(b) we show the time dependence of the average cluster size. One can clearly see the 
presence of two distinct phases. The early-time phase corresponds to the stage of stable cluster 
formation ($\le 10^6 $ MCS) and the later phase is the coarsening phase. The early-time data are consistent 
with a behavior $C_s(t) \sim t^{1/4}$ which is slower than the corresponding behavior in $d=3$ (see Fig.\ 8(b) in Ref.\ \cite{majumder2017SM}). 
The late-time behavior is consistent with a $C_s(t) \sim t$ behavior consistent with a $d=3$ polymer using OLM. 
However, we caution the reader that one must be careful before interpreting the linear behavior. In this regard, 
we believe that a proper finite-size scaling analysis as done for the $d=3$ case is required to confirm it, for which 
one needs data from different system sizes. This analysis is in progress and will be presented elsewhere. 

\subsection{Aging in $d=2$}
\begin{figure}[t!]
\centering
\resizebox{0.9\columnwidth}{!}{\includegraphics{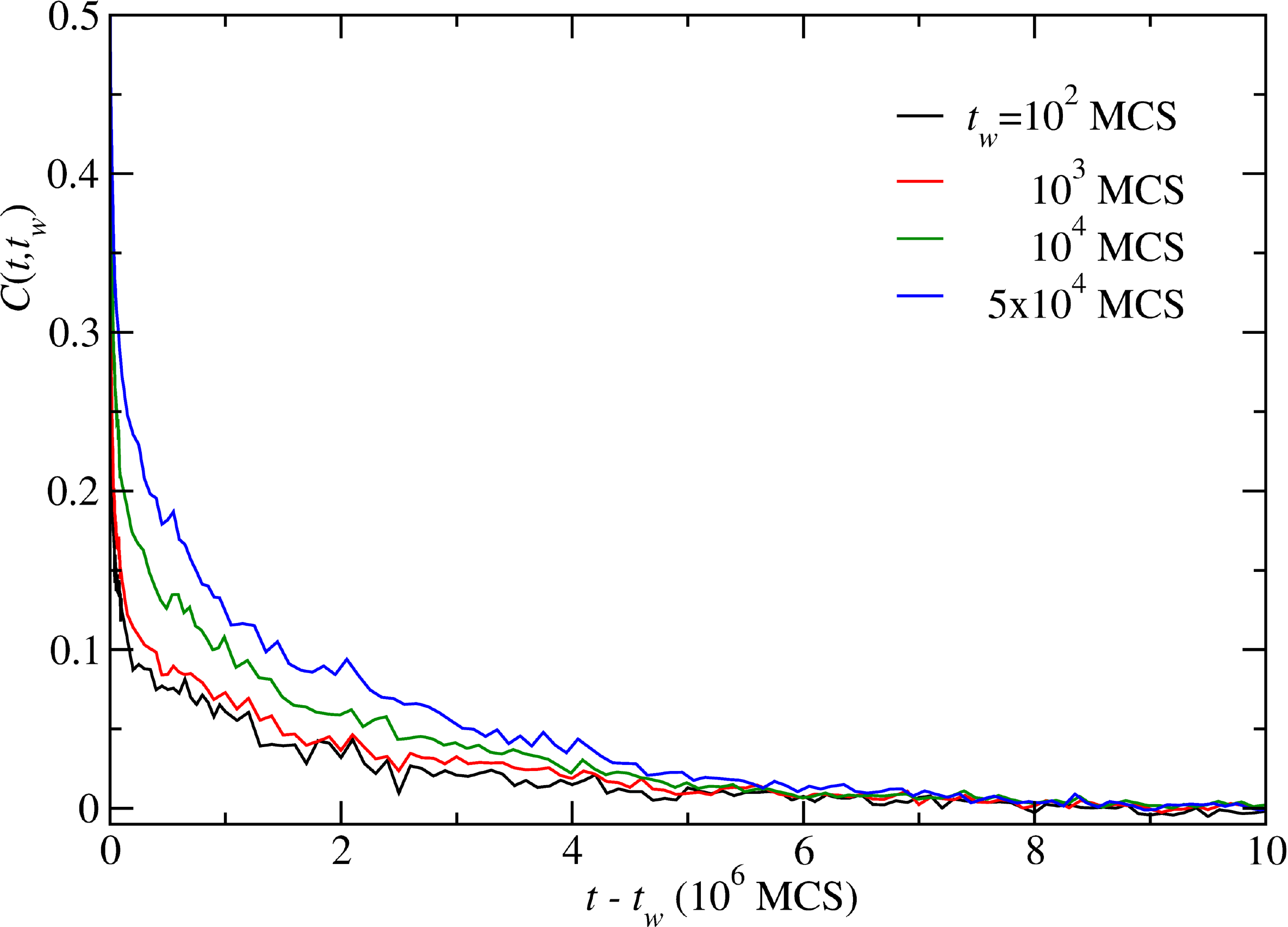}}
\caption{Demonstration of the breakdown of time-translation invariance by plotting the autocorrelation function 
$C(t,t_w)$ as a function of the translated time $t-t_w$, during collapse of a polymer in $d=2$ modeled 
by the OLM. The chain length and $T_q$ are the same as in Fig.\ \ref{noc2d}. The chosen values of the waiting times $t_w$
are mentioned within the graph.}
\label{corrtmtw}      
\end{figure}

\begin{figure}[b!]
\centering
\resizebox{0.9\columnwidth}{!}{\includegraphics{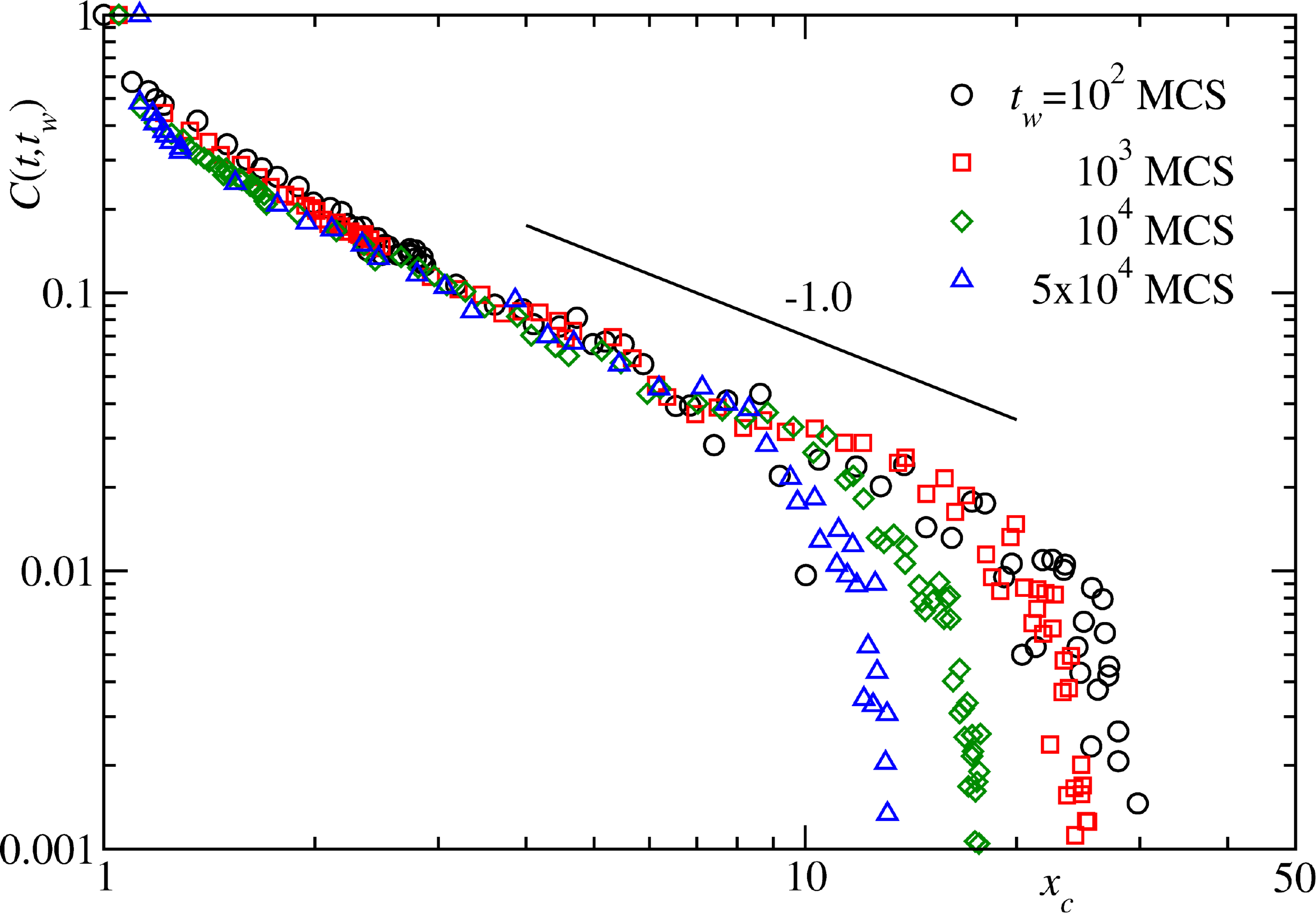}}
\caption{Illustration of the presence of dynamical scaling of the autocorrelation function shown in Fig.\ \ref{corrtmtw}, plotted here on a double-log scale 
as a function of the scaling variable $x_c=C_s(t)/C_s(t_w)$. The solid line shows the consistency of the data with a power-law decay having 
an exponent $\lambda_c=1.0$.}
\label{corrlbylw}      
\end{figure}

We now move on to present some preliminary results on the aging dynamics during polymer collapse in $d=2$ using the OLM. 
Like in the $d=3$ case here also, we probe aging via calculation of the two-time autocorrelation function described in \eqref{auto_cor} by using 
the same criterion for $O_i$ as used in $d=3$ for the OLM. To check the presence of aging we first confirm 
the absence of time-translation invariance. This is demonstrated in Fig.\ \ref{corrtmtw} for the same system as presented for the 
cluster growth in Fig.\ \ref{noc2d}. The plot shows the autocorrelation function $C(t,t_w)$ as a function of the translated time 
$t-t_w$ for four different values of $t_w$ as mentioned in the figure. The absence of time-translation invariance is evident from the non-collapsing 
behavior of the data. Along with that one can also notice that the larger $t_w$ the slower the autocorrelation decays which confirms 
the second criterion of aging, i.e., slow dynamics. The last criterion for aging is the presence of dynamical scaling. 
In the present case of polymer collapse in $d=2$, unlike in the $d=3$ case with OLM, we do not observed any data collapse with 
respect to the scaling variable $x_c=t/t_w$. This, on the other hand, is similar to the results obtained for the LM in $d=3$. 
However, to limit ourselves here rather than going for an analysis based on subaging scaling we immediately 
look for the scaling with respect to $x_c=C_s(t)/C_s(t_w)$ and indeed find a reasonable collapse of data 
implying the presence of simple aging behavior. This is demonstrated in Fig.\ \ref{corrlbylw} where we plot $C(t,t_w)$ 
as a function of $x_c=C_s(t)/C_s(t_w)$ for four different choices of $t_w$.
\par
The other important aspect of aging is to quantify the autocorrelation exponent $\lambda_c$ for which an idea can be obtained from the double-log plot in Fig.\ \ref{corrlbylw}. There 
for intermediate values of $x_c$, the collapsed data show almost a linear behavior implying a power-law scaling. The solid line corresponds to the power-law decay in Eq.\ \eqref{power-law_Cst}
with an exponent $\lambda_c=1$ that is consistent with the data. For a better quantification of 
$\lambda_c$ one would need to do a finite-size scaling analysis by using data from a few larger chain lengths. From the general bound given 
in Eq.\ \eqref{poly-bound}, one can read off the corresponding bound in $d=2$, 
\begin{eqnarray}\label{2d-bound}
 0.5 \le \lambda_c \le 1.0,
 \end{eqnarray}
where we have used the fact that in $d=2$, the Flory exponent is exactly $\nu_F=0.75$ \cite{Florybook,vanderzandebook}. The consistency of our data in Fig.\ \ref{corrlbylw} with the autocorrelation exponent 
$\lambda_c=1$ implies that in $d=2$ the bound is marginally obeyed. However, to have an appropriate verification of the bound one needs to have a more reliable estimate of $\lambda_c$ as already mentioned. 

\section{Conclusion and outlook}\label{conclusion}
We have presented an overview of results existing in the literature regarding the 
collapse dynamics of a homopolymer. Although research in this direction started long back with the 
proposition of the sausage model of collapse by de Gennes, after a series of works by Dawson and co-workers \cite{byrne1995,timoshenko1995,kuznetsov1995,kuznetsov1996,kuznetsov1996eDNA,dawson1997}
and a few other \cite{pitard1998,klushin1998,Halperin2000,kikuchi2002,Abrams2002,Montesi2004,yeomans2005}, it eventually faded away. Particularly, in experiments it was difficult to monitor a single 
polymers to verify the phenomenological theories developed around collapse dynamics. Recently, motivated 
by the successful experimental development for monitoring single polymers and polymers in very dilute solutions, 
we have provided some new insights in the collapse dynamics of polymers via computer simulations. In this regard, we borrowed 
tools and understanding from the general nonequilibrium process of coarsening in particle and spin systems. 
This allowed us to explore different nonequilibrium scaling laws that could be associated with 
kinetics of the collapse transition of polymers.

\par
When speaking of scaling laws concerning collapse dynamics of a polymer the first thing one looks 
for is the scaling of the overall collapse time $\tau_c$ with the chain length $N$  (which was also the main focus of the studies in the past). From a survey of the available results in this direction it is clear that for 
power-law\- scaling of the form $\tau_c \sim N^z$, the value of the dynamical exponent $z$ obtained 
depends on the intrinsic dynamics used in the simulations. Especially one has to be careful about 
presence of hydrodynamics while quoting the value of $z$. However, in our work with 
an off-lattice model via Monte Carlo dynamics for large $N$, we obtained a value of $z$ that is close 
to the one obtained from molecular dynamics simulations with preservation of hydrodynamic effects. This 
raises the question of to what extent hydrodynamics interactions are important during collapse. A proper answer to this could 
be obtained via systematic studies of polymer models with explicit solvent \cite{pham2008,chang2001solvent,polson2002}. For the latter there also exist few studies; however,
with no consensus about the value of $z$. In the context of 
doing simulations with explicit solvent it would also be interesting to see the effect of 
the viscosity of the solvent particles on the dynamics. Building of such a framework is 
possible with an approach based on the dissipative particle dynamics \cite{hoogerbrugge1992,espanol1995,groot1997,espanol2017}. Recently, we have taken up 
this task by using an alternative approach to dissipative particle dynamics \cite{lowe1999,koopman2006}. 
In this context, we have successfully constructed the set up and tested that it reproduces the correct dynamics 
in equilibrium taking consideration of the hydrodynamic interactions appropriately \cite{majumder2019dissipative}. 
To add more to this understanding recently we have also considered the task of doing all-atom 
molecular dynamics simulations with explicit solvent \cite{majumder2019macro}. There the focus is on understanding the collapse of a polypeptide in water 
with the aim to get new insights to the overall folding process of a protein which 
contains these polypeptides as backbone. 
\par
Coming back to the scaling laws during collapse our approach of understanding the collapse 
in analogy with usual coarsening phenomena allows us to explore the cluster kinetics 
appropriately. Our findings from studies using both off-lattice and lattice models 
show that the average cluster size $C_s(t)$ during the collapse grows in a power-law fashion 
as $C_s(t) \sim t^{\alpha_c}$. However, the growth exponent $\alpha_c$ is not universal 
with $\alpha_c \approx 1$ for the off-lattice model and $\alpha_c \approx 0.62$ for the lattice model. 
For quantification of this growth exponent one must be careful about the initial cluster formation stage 
which sets a high off-set while fitting the data to a simple power law. In this regard, we have 
introduced a nonequilibrium finite-size scaling analysis which helps to estimate 
the value of $\alpha_c$ unambiguously.

\par
Along with the growth kinetics where one deals with single-time quantities, it is also important 
to have understanding of the multiple-time quantities which provide information 
about the aging during such nonequilibrium processes. In analogy with the two-time density or order-parameter 
autocorrelation function used in usual coarsening of particle or spins systems, we have shown how one can 
construct autocorrelation functions to study aging during collapse of a polymer. Depending on the 
nature of the model (whether off-lattice or lattice) the chosen observable to calculate the 
autocorrelation may vary; however, qualitatively they should give the same information. Our results 
indeed support our choice of the respective observables and provide evidence of aging and corresponding dynamical scaling 
of the form $C(t,t_w) \sim \left[C_s(t)/C_s(t_w)\right ]^{-\lambda_c}$. Unlike the growth exponent, the 
dynamic aging exponent was found to be $\lambda_c=1.25$ irrespective of the nature of the model, implying that 
the aging behavior is rather universal. In this regard, it is worth mentioning that even choosing two different 
bond criteria for the lattice model (one with the diagonal bonds and the other without it \cite{christiansen2017JCP}) 
yielded cluster growth exponents that are different, however, the aging 
exponent $\lambda_c$ still remains universal with a value of $1.25$. To check the robustness of this universality, 
a study of other polymer models both off-lattice and lattice, along with different methods of 
simulations as mentioned previously is required. 

\par
In addition to the review of the existing results we have also presented preliminary results 
in the context of polymer collapse in $d=2$ dimensions. To understand a two-dimensional system is not only of fundamental 
interest \cite{Jia_PRL}, but could be of relevance in the context of polymers confined to an attractive surface. Indeed 
there are experiments of synthetic polymers on two-dimensional gold or silver surfaces \cite{forster2014structure,forster2014}. Our results 
on the kinetics of polymer collapse in $d=2$ show that the phenomenology 
associated with this process can still be described by the ``pearl-necklace'' picture of Halperin and Goldbart, albeit 
the identification of the small pearl-like clusters which coarsen to form the final globule 
is not as distinct as in the $d=3$ case. Via an extension of the $d=3$ methodologies to $d=2$ , we observe that the 
cluster formation stage in $d=2$ is rather slow. However, the late-time coarsening of the clusters 
follows the same power-law scaling $C_s(t) \sim t^{\alpha_c}$ with $\alpha_c \approx 1$. 
We also have presented results for the aging dynamics in this regard as well. There the autocorrelation function 
shows the same kind of power-law scaling as in $d=3$ with a corresponding exponent $\lambda_c\approx 1$. 
A more detailed study not only with the off-lattice model but also with the lattice 
model is in progress. 

\par
Finally, we feel that this novel approach of understanding the collapse dynamics of polymers from 
the perspective of usual coarsening studies of particle and spin systems shall serve as a general platform which could 
be used to analyze the nonequilibrium evolution of macromolecules in general across any conformational transition. 
Of course, due to their distinct features, for each class of this transition the associated techniques 
shall be modified accordingly. One has to choose the appropriate properties of the system 
and find out the best quantities that describe the corresponding transition appropriately in nonequilibrium. 
For example, one can look at the helix-coil transition of macromolecules as well \cite{Arashiro2006,Arashiro2007}. There certainly the 
average cluster size would not work as a suitable quantity to monitor the kinetics. Rather one may define some 
local helical order parameter and look at the corresponding time dependence.
\begin{acknowledgments}
This project was funded by the Deutsche Forschungsgemeinschaft (DFG, German Research Foundation) 
under project Nos. JA 483/33-1 and 189\,853\,844 -- SFB/TRR 102 (project B04), and the 
Deutsch-Franz\"osische Hoch\-schule (DFH-UFA) through the Doctoral College ``$\mathbb{L}^4$'' 
under Grant No.\ CDFA-02-07. We further acknowledge support by the Leipzig Graduate School of Natural Sciences ``BuildMoNa''.
\end{acknowledgments}
\section*{Author contribution statement}
S.M. planned the structure of the manuscript with inputs from the co-authors. All the authors 
contributed equally in writing and developing the text. 

%
%
\bibliography{bib_new}

\end{document}